%% file: great08_results.tex
\documentclass[useAMS,usenatbib]{mn2e}
\usepackage{amsmath,amssymb,epsfig,natbib,bm,psfrag,ifthen,url}
\usepackage{hyperref}
\usepackage{tikz}
\usetikzlibrary{arrows}
\usetikzlibrary{er}

\begin{document}

\newcommand{\mnras}{Monthly Notices of the Royal Astronomical Society}
\newcommand{\aj}{Astronomical Journal}
\newcommand{\apj}{Astrophysical Journal}
\newcommand{\apjl}{Astrophysical Journal Letters}
\newcommand{\aap}{Astronomy and Astrophysics}
\newcommand{\araa}{Annual Review of Astronomy and Astrophysics}

\def\eprinttmp@#1arXiv:#2 [#3]#4@{
\ifthenelse{\equal{#3}{x}}{\href{http://arxiv.org/abs/#1}{#1}
}{\href{http://arxiv.org/abs/#2}{arXiv:#2} [#3]}}

\providecommand{\eprint}[1]{\eprinttmp@#1arXiv: [x]@}
\newcommand{\adsurl}[1]{\href{#1}{ADS}}

\title[Results of the GREAT08 Challenge]
{Results of the GREAT08 Challenge\thanks{\url{http://www.great08challenge.info}}: An image analysis
  competition for cosmological lensing \vspace{-0.8cm}}
\author[S. L. Bridle et al.]
{Sarah Bridle$^{1}$\thanks{E-mail: sarah.bridle@ucl.ac.uk},
Sreekumar T. Balan$^{2}$, Matthias Bethge$^{3}$, Marc Gentile$^{4}$,
\newauthor
Stefan Harmeling$^{3}$, Catherine Heymans$^{5}$, Michael Hirsch$^{3}$, Reshad Hosseini$^{3}$,
\newauthor
Mike Jarvis$^{6}$, Donnacha Kirk$^{1}$, Thomas Kitching$^{5}$, Konrad Kuijken$^{7}$,
\newauthor
Antony Lewis$^8$, Stephane Paulin-Henriksson$^{9}$, Bernhard Sch\"{o}lkopf$^{3}$,
\newauthor
Malin Velander$^{7}$, Lisa Voigt$^{1}$, Dugan Witherick$^{1}$,
Adam Amara$^{10}$, Gary Bernstein$^6$,
\newauthor
Fr\'ed\'eric Courbin$^{4}$, Mandeep Gill$^{11}$, Alan Heavens$^5$,
Rachel Mandelbaum$^{12}$,
\newauthor
Richard Massey$^{5}$, Baback Moghaddam$^{13,14}$,
Anais Rassat$^{9}$,
\newauthor
Alexandre R\'efr\'egier$^{9}$, Jason Rhodes$^{13,14}$, Tim Schrabback$^{7}$,
John Shawe-Taylor$^1$,
\newauthor
Marina Shmakova$^{15}$, Ludovic van Waerbeke$^{16}$, David Wittman$^{17}$
\\
$^{1}$Department of Physics and Astronomy, University College London, Gower Street, London, WC1E 6BT, UK.\\
$^{2}$Cavendish Astrophysics, University of Cambridge, JJ Thomson Avenue, Cambridge CB3 0HE, UK.\\
$^{3}$MPI for Biological Cybernetics, Dept. of Empirical Inference, Spemannstrasse 38, 72076 T\"{u}bingen, Germany.\\
$^{4}$ Ecole Polytechnique F\'{e}d\'{e}rale de Lausanne (EPFL), Observatoire de Sauverny, 1290 Versoix, Switzerland.\\
$^{5}$Institute for Astronomy, University of Edinburgh, Royal Observatory, Blackford Hill, Edinburgh, EH9 3HJ, UK.\\
$^{6}$Department of Physics and Astronomy, University of Pennsylvania, Philadelphia, PA 19104, USA.\\
$^{7}$Leiden Observatory, P.O. Box 9513, NL-2300 RA, Leiden, The Netherlands.\\
$^{8}$Institute of Astronomy and Kavli Institute for Cosmology, Madingley Road, Cambridge, CB3 0HA, UK\\
$^{9}$Service dAstrophysique, CEA Saclay, F-91191 Gif sur Yvette, France.\\
$^{10}$Department of Physics, ETH Z\"{u}rich, Wolfgang-Pauli-Strasse 16, CH-8093 Z\"{u}rich, Switzerland.\\
$^{11}$Department of Astronomy, Ohio State University, 140 W. 18th Avenue, Columbus, OH 43210, USA.\\
$^{12}$Institute for Advanced Study, Einstein Drive, Princeton, NJ 08540, USA.\\
$^{13}$Jet Propulsion Laboratory, California Institute of Technology, 3800 Oak Grove Drive, Pasadena, A 91109, USA.\\
$^{14}$California Institute of Technology, Pasadena, CA 91125, USA.\\
$^{15}$Stanford Linear Accelerator Center, Stanford University, P.O. Box 4349, CA 94309, USA.\\
$^{16}$University of British Columbia, 6224 Agricultural Rd., Vancouver, BC, V6T 1Z1, Canada.\\
$^{17}$Department of Physics, University of California at Davis, One Shields Avenue, Davis, CA 95616, USA.
\vspace{-0.5cm}
}

\maketitle

\label{firstpage}

\begin{abstract}
We present the results of the GREAT08 Challenge, a blind analysis challenge to infer weak gravitational lensing shear distortions from images. The primary goal was to stimulate new ideas by presenting the problem to researchers outside the shear measurement community. Six GREAT08 Team methods were presented at the launch of the Challenge and five additional groups submitted results during the 6 month competition. Participants analyzed 30 million simulated galaxies with a range in signal to noise ratio, point-spread function ellipticity, galaxy size, and galaxy type. The large quantity of simulations allowed shear measurement methods to be assessed at a level of accuracy suitable for currently planned future cosmic shear observations for the first time. Different methods perform well in different parts of simulation parameter space and come close to the target level of accuracy in several of these. A number of fresh ideas have emerged as a result of the Challenge including a re-examination of the process of combining information from different galaxies, which reduces the dependence on realistic galaxy modelling. The image simulations will become increasingly sophisticated in future GREAT challenges, meanwhile the GREAT08 simulations remain as a benchmark for additional developments in shear measurement algorithms.
\end{abstract}

\begin{keywords}
cosmology: observations - gravitational lensing - large-scale structure
\end{keywords}

\section{Introduction}

A clump of matter induces a curvature in space-time which causes the trajectory of a light ray to appear bent. This effect, known as gravitational lensing, is analogous to light passing through a sheet of glass of varying thickness such as a bathroom window. In both cases the light-emitting objects appear distorted. Making assumptions about the intrinsic (original) shapes of the emitting objects allows us to infer information about the intervening material. In cosmology we learn about the distribution of matter by studying the shapes of distant galaxies. In the vast majority of cases the distortion varies very little as a function of position on the galaxy image, and it can be approximated by a matrix distortion. This regime is known as weak gravitational lensing, or cosmic shear when applied to large numbers of randomly selected distant galaxies.

Gravitational attraction of ordinary matter and dark matter is expected to slow the expansion of the universe, causing the expansion to decelerate. However, multiple lines of evidence now show that the present day expansion of the Universe seems instead to be accelerating. The main explanations explored in the literature are that (i) Einstein's cosmological constant is non-zero, (ii) the vacuum energy is small but non-negligible, (iii) the Universe is filled with some new fluid, dubbed dark energy, or (iv) the laws of General Relativity are wrong at large distances. Possibilities (i) and (ii) can be subsumed within item (iii) because they look like a dark energy fluid with equation of state $p=w\rho c^2$ where $w=-1$. To find out more about the nature of dark energy or modifications to the law of gravity we need high precision measurements of the recent $(z<1)$ Universe.

By studying cosmic shear using galaxies at a range of different epochs we can learn how the dark matter clumps as a function of time, which itself depends on the nature of dark energy and the laws of gravity. Cosmic shear appears to hold the most potential of all methods for investigating the dark energy or modifications to gravity~\citep{detf,Peacock:2006kj,albrechtb07,FOMSWG}. There are many current, planned and proposed surveys to use cosmic shear to measure dark energy including the Canada-France Hawaii Telescope Legacy Survey (CFHTLS)~\footnote{\url{http://www.cfht.hawaii.edu/Science/CFHLS/}}, the KIlo-Degree Survey (KIDS), Panoramic Survey Telescope and Rapid Response System (Pan-STARRS)~\footnote{\url{http://pan-starrs.ifa.hawaii.edu}}, the Dark Energy Survey (DES)~\footnote{\url{http://www.darkenergysurvey.org}}, the Large Synoptic Survey Telescope (LSST)~\footnote{\url{http://www.lsst.org}}, and space missions Euclid~\footnote{\url{http://sci.esa.int/euclid}} and the Joint Dark Energy Mission (JDEM)~\footnote{\url{http://jdem.gsfc.nasa.gov}}.

Cosmic shear was first detected just one decade ago \citep{baconre00,Kaiser:2000if,van_Waerbeke:2000rm,Wittman:2000tc} and many studies have now used it to measure cosmological parameters. Much work has also been carried out on anticipating any problems that may limit the potential of cosmic shear over the coming decade. These are thought to be (i) accuracy of approximate methods for obtaining distances to galaxies; (ii) intrinsic alignments of galaxies; (iii) accuracy of numerical predictions of dark matter clustering on small scales and in the presence of baryons; and (iv) unbiased measurement of shear from galaxy images. There is now much discussion about obtaining high quality galaxy distances using spectroscopic redshifts to calibrate approximate methods to solve (i) \citep{mahh05,huterertbj06,kitchingth08,bernsteinm08,bernsteinh09}. The intrinsic alignment signal (ii) can be removed if (i) can be solved perfectly \citep{takadaw04,joachimis08} and otherwise the two are closely linked \citep{kings02b,heymansh03,king05,bridlek07,zhang08,bernstein09,joachimis09}. Supercomputers are being deployed to produce higher accuracy predictions, and methods for suppressing information from the uncertain small-scale regime have been developed. In this paper we focus on the final problem, shear measurement from noisy images. It can be phrased entirely as a statistics problem of extracting information from images.

In 2004 the Shear TEsting Programme (STEP) was launched to assess the current status of shear measurement methods. It began with a blind challenge set by and for the weak lensing community~\citep[][hereafter STEP1]{STEP1mnras}. A large volume of images containing a mixture of stars and simple galaxies were produced. The participants had the task of extracting the (constant) input shear from the images, and these estimates were compared to the true input value. These end-to-end simulations showed that the shear measurement problem is far from trivial but that the methods in frequent use at that time were sufficiently accurate for the existing published cosmic shear measurements. \cite{STEP2mnras} (hereafter STEP2) extended this work with more sophisticated galaxy models, and built statistical devices into larger simulations to improve the measurement precision. This showed that, even considering realistic and more complex galaxy morphologies, existing methods were still sufficient for the current data.

The cosmic shear community then began to look ahead to the coming decade of surveys and ask whether the existing methods are sufficiently accurate even when the statistical uncertainties are reduced by the massive increase in data quantity. Addressing this question requires much larger blind challenges, containing at least tens of millions of galaxies. At the same time it was recognised that the shear estimation problem can be phrased as a statistics problem and that experts in image analysis from other disciplines may be in a position to contribute significantly to developing new approaches. Furthermore, it was decided that the strengths and weaknesses of different methods could be best assessed with slightly simpler simulations, in which various effects could be isolated.

The previous two published blind shear analysis challenges (STEP1, STEP2) were slightly simplified relative to real data in that the shear and the PSF did not vary across an image.  However, they did ask participants to grapple with a number of difficult issues.
\\$\bullet$ The images had relatively realistic PSFs with classical optical
  aberrations such as coma and trefoil.
\\$\bullet$ Although the PSF did not vary across an image, participants were
  asked not to use this fact.
\\$\bullet$ STEP1 required participants to determine which objects were
  stars and therefore could be used for a PSF determination.
\\$\bullet$ Both challenges required participants to run object detection
  software to determine where the star and galaxies were.  Spuriously
  detected objects could and did affect the shear.
\\$\bullet$ Galaxies were drawn from a range of magnitudes, so that
  weighting schemes
  as a function of the Signal-to-Noise Ratio (SNR)
  were
  important.
\\$\bullet$ Galaxies were randomly placed, so that sometimes they
  overlapped. Participants were responsible for either deblending or
  rejecting these galaxies.
\\
The GREAT08 Challenge removes {\it all} of these issues to focus on the core problem of inferring shear given a PSF and standardised set of non-overlapping galaxies at (approximately) known positions.  The motivation is that once this problem is solved, the other issues will be introduced in further challenges of increasing complexity.

The Gravitational LEnsing Accuracy Testing 2008 (GREAT08) Challenge Handbook \citep[][hereafter The GREAT08 Handbook]{great08mnras} describes the shear measurement problem for non-cosmologists and sets out the challenge. GREAT08 was launched in October 2008 and ran as a blind competition for 6 months until the end of April 2009. This paper describes the results of GREAT08. Section~\ref{sect:sims} describes the GREAT08 simulations. We review the shear measurement problem and shear accuracy requirements in Section~\ref{sect:Requirements}. Section~\ref{sect:Methods} summarises current shear measurement methods and Section~\ref{sect:Results} presents the Challenge results. We conclude and overview the potential for future GREAT Challenges in Section~\ref{sect:discussion}. We provide extra details of the simulations, methods and results in appendices.

\section[]{The GREAT08 Simulations}
\label{sect:sims}

\begin{figure*}
\center
\epsfig{file=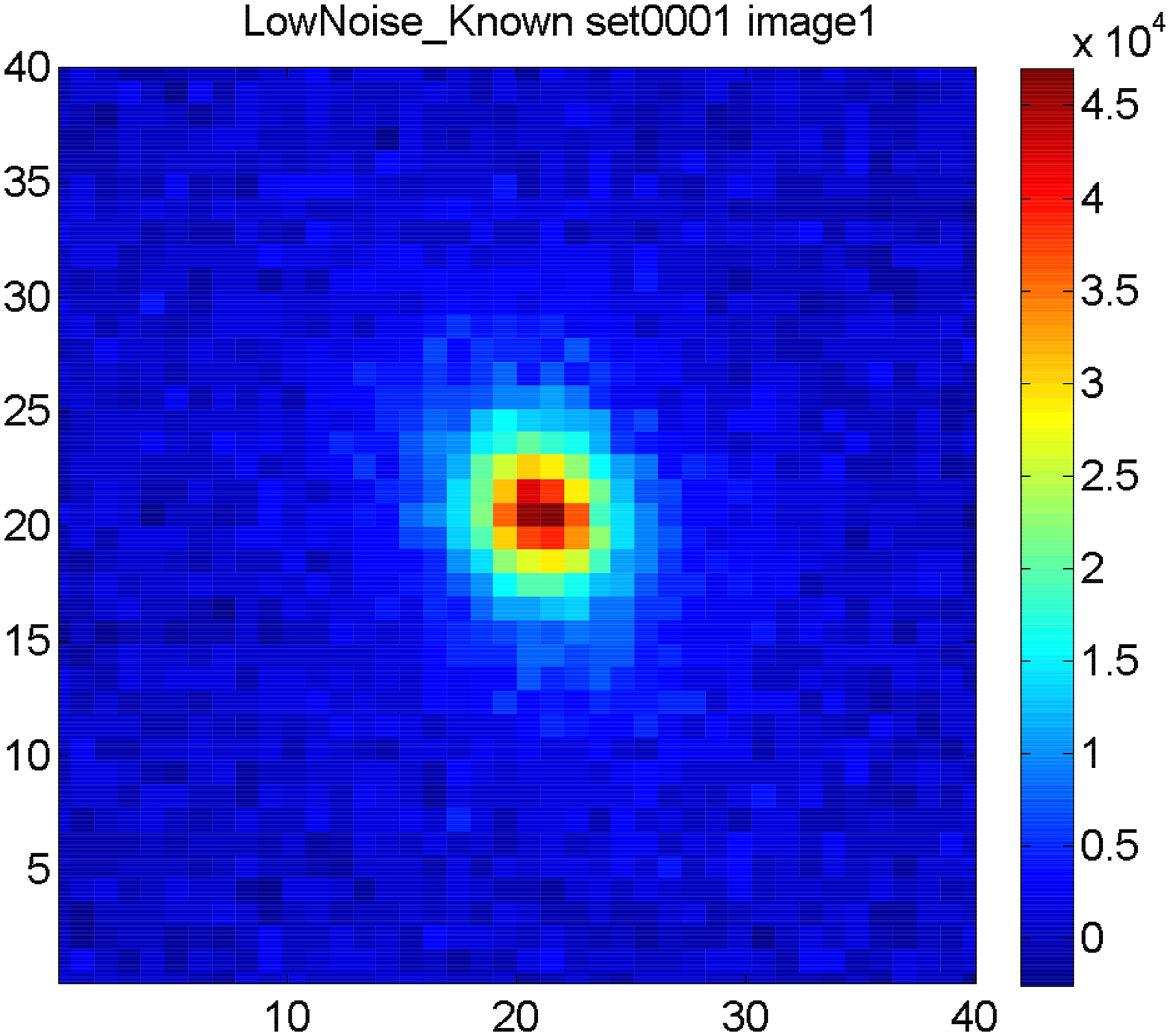,width=8cm,angle=0}
\epsfig{file=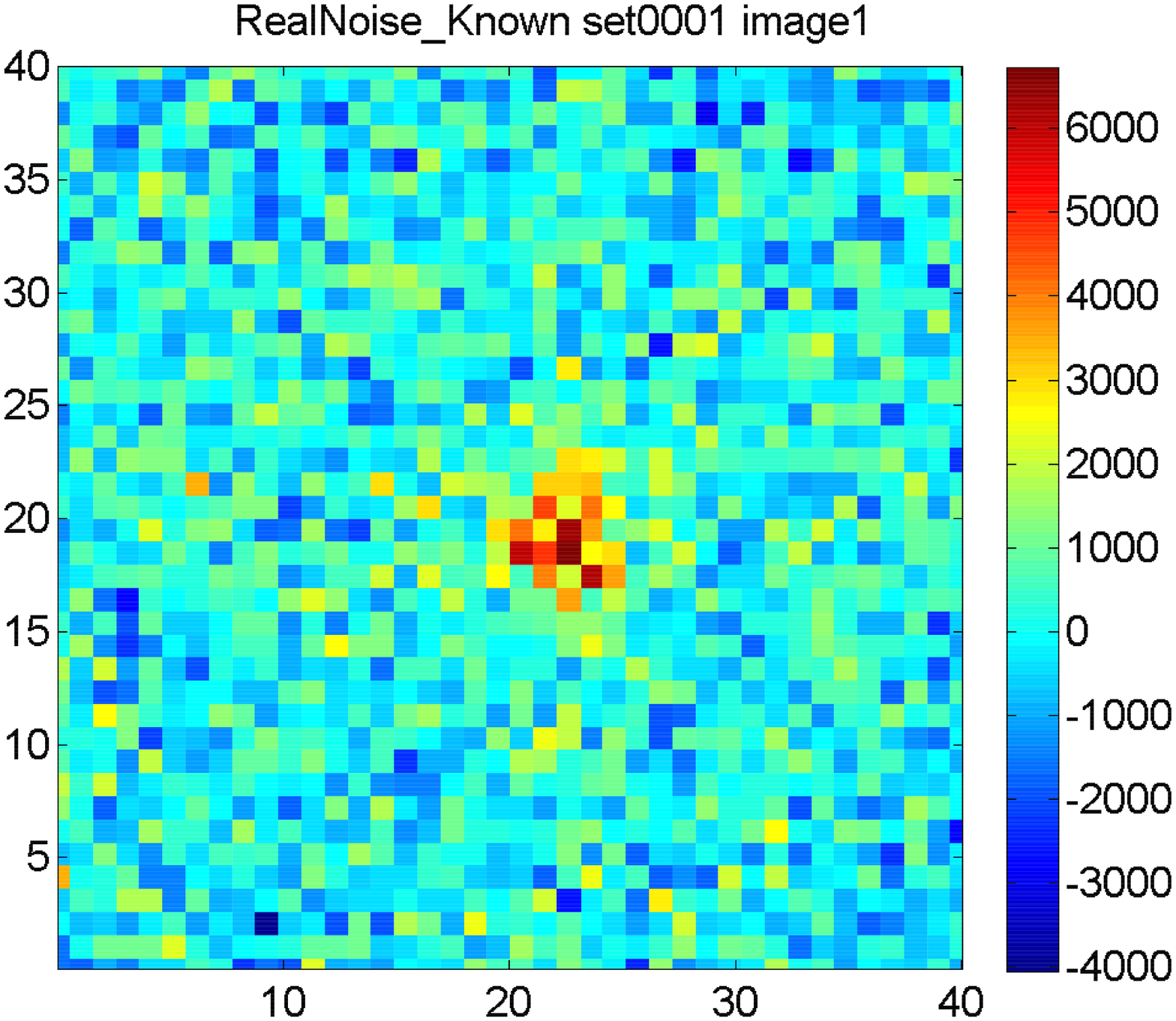,width=8cm,angle=0}
\caption{
Left: The first galaxy of the first LowNoise\_Known {\tt FITS} image.
Right: The first galaxy of the first RealNoise\_Known {\tt FITS} image.
The signal is a factor of ten smaller for the RealNoise images than the LowNoise images, making the problem much more challenging.
}
\label{fig:images}
\end{figure*}

The GREAT08 images are provided in sets of 10,000 objects in a single {\tt FITS} file. Each object is generated on its own grid of $39\times39$ pixels and these postage stamps are patched together for convenience in a $100\times100$ layout, with a 1 pixel border, thus each set is a patchwork image of 4000x4000 pixels.
Each galaxy postage stamp is generated using the following sequence:
(i) simulate a galaxy model; (ii) convolve it with a kernel, referred
to as the point-spread function (PSF); (iii) bin up the light in
pixels; and (iv) apply the noise model.  The PSFs used are given in
Appendix~\ref{appendix:PSF_models}.  Each postage stamp is produced using
a list of parameters specifying the individual object and simulation
properties. We describe the catalogues of these properties in
Appendix~\ref{appendix:Galaxy_catalogue_generation}.  The method used to
produce images from the catalogues is overviewed below and described in more detail in
Appendix~\ref{appendix:Image_simulations}. Example images are shown in Fig.~\ref{fig:images}.

\begin{figure*}
\center
\epsfig{file=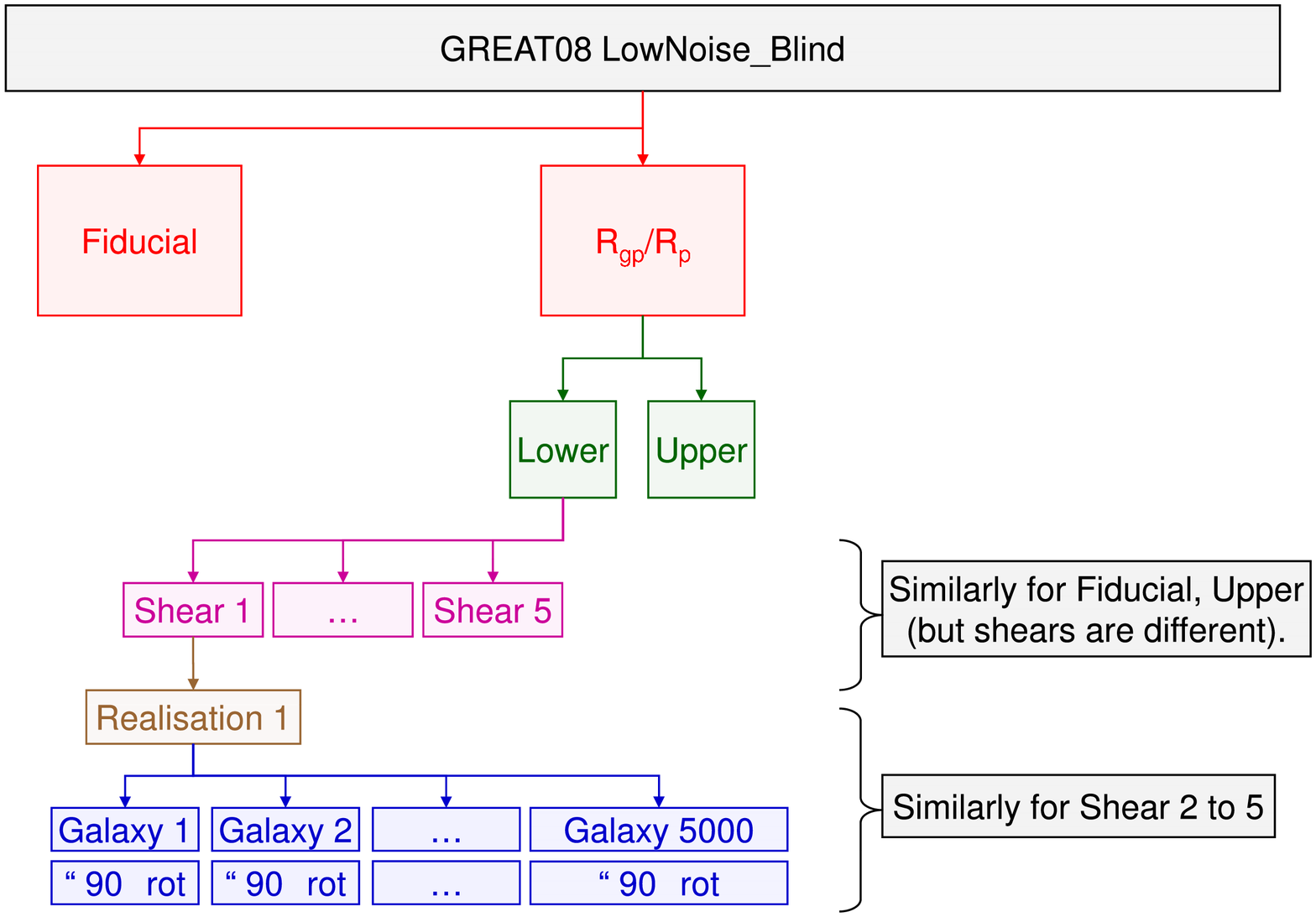,width=15cm,angle=0}
\epsfig{file=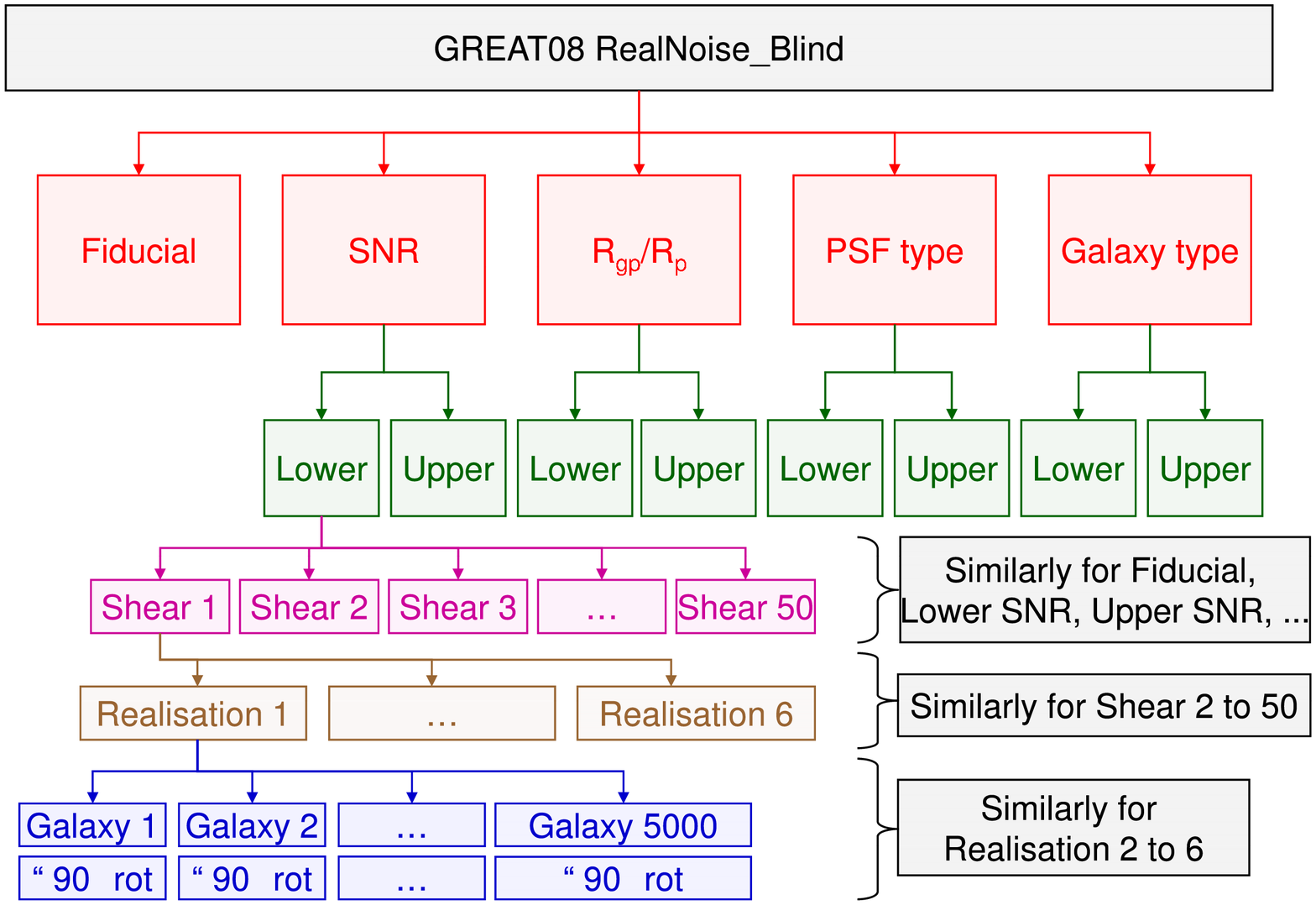,width=15cm,angle=0}
\caption{
{\it Upper panel:} Schematic of the galaxy parameters used in LowNoise$\_$Blind. Each realisation corresponds to a different set or {\tt FITS} image file containing 10,000 galaxies.
The schematic looks identical for LowNoise$\_$Known. For RealNoise$\_$Known there are 100 shears per branch in place of 5.
The bottom row of boxes represents galaxies with the same properties
as the penultimate row of boxes, but rotated by 90 degrees.
{\it Lower panel:}
Schematic of the galaxy parameters used in RealNoise$\_$Blind.
 }
\label{fig:tree}
\end{figure*}

Four different groups of galaxy images were provided in GREAT08: (i) low noise galaxy images for which the true shears were provided during the Challenge, labelled LowNoise$\_$Known; (ii) low noise galaxy images for which there was a blind challenge to extract the true shears, labelled LowNoise$\_$Blind; (iii) realistic noise galaxy images for which the true shears were provided, labelled RealNoise$\_$Known; and (iv) realistic noise galaxy images with blind shear values, RealNoise$\_$Blind. This RealNoise$\_$Blind group formed the main GREAT08 Challenge. These are described in more detail in the GREAT08 Handbook, together with the rules governing which information could be used to inform the blind challenges.

\begin{table}
\caption{
Parameters for the LowNoise$\_$Known simulations.
$R_{gp}/R_p$ is the ratio of PSF convolved galaxy
Full Width at Half Maximum (FWHM)
to the PSF FWHM.
`b or d' describes the fact that
50\% of the galaxies in each set have de Vaucouleurs profiles (bulge only) and 50\% have exponential profiles (disk only). The parameters for LowNoise$\_$Blind are the same except the galaxies are a mix of the two components as described in the text. The parameters for RealNoise$\_$Known are the same as for LowNoise$\_$Known except the SNR is 20.
}
\label{tab:LowNoise_Known_pars}
\center{
\begin{tabular}{|c|c|c|c|}
\hline
& Fiducial & Lower value & Upper value\\
\hline
SNR & 200 & N/A & N/A\\
$R_{gp}/R_p$ & 1.4 & 1.22 & 1.6\\
PSF type        & Fid & N/A & N/A\\
Galaxy type     & b or d & N/A & N/A\\
\hline
\end{tabular}
}
\end{table}

\begin{table}
\caption{
Parameters for the RealNoise$\_$Blind simulations.
The PSF models and other parameters are defined in detail in Appendices~\ref{appendix:PSF_models} and~\ref{appendix:Galaxy_catalogue_generation}.
}
\label{tab:RealNoise_Blind_pars}
\center{
\begin{tabular}{|c|c|c|c|}
\hline
& Fiducial & Lower value & Upper value\\
\hline
SNR & 20 & 10 & 40 \\
$R_{gp}/R_p$ & 1.4 & 1.22 & 1.6\\
PSF type        & Fid & Fid rotated & Fid $ e \times$ 2\\
Galaxy type     & b+d & b or d & b+d offcenter\\
\hline
\end{tabular}
}
\end{table}

The parameters for each set in LowNoise$\_$Known were determined using
the upper panel of
Fig.~\ref{fig:tree} and Table~\ref{tab:LowNoise_Known_pars}.
There are 15 sets ({\tt FITS} images) each containing 10,000 galaxies.
There are 5 sets with each of 3 different galaxy size values.
The method for setting the galaxy sizes and SNR values is described
in Appendices~\ref{appendix:Galaxy_catalogue_generation} and~\ref{appendix:Image_simulations}.

The parameters for each set in RealNoise$\_$Blind were determined using
the lower panel of
Fig.~\ref{fig:tree} and Table~\ref{tab:RealNoise_Blind_pars}.
There is a range in SNR, galaxy size, PSF ellipticity and galaxy type.
One branch of the RealNoise$\_$Blind holds all parameters at their fiducial values. Each of the 4 variable parameters has a `lower' and an `upper' value relative to the fiducial. When each of these values is used all other parameters are fixed at the fiducial values. This makes 9 different branches in total. In each branch there are 6 realisations of each of 50 different shear values, making 2700 sets with 10,000 galaxies in each.

Images are generated by sampling from the galaxy light distribution, sampling from the PSF, adding the sample positions to simulate convolution, binning the samples onto a pixel grid, and then applying the noise model. The exact numerical techniques used are detailed in Appendix~\ref{appendix:Image_simulations}.
In brief, samples are first generated from the circular galaxy profile.
Next, they are stretched to have the required ellipticity and then sheared.
Samples are then drawn from the circular PSF distribution and made elliptical using
the shear distortion equations given in Appendix~\ref{appendix:Image_simulations}.
Each galaxy sample is added to a PSF sample to simulate convolution, and finally the samples are binned into pixels.

\section[]{Figure of Merit}
\label{sect:Requirements}

The shear measurement problem was summarised for non-cosmologists in
the GREAT08 Handbook. In short, light from a source galaxy is sheared
and (slightly) magnified by passing through a gravitational potential
on its way to the observer; the observable anisotropic stretching is
called the {\it reduced shear} $g$, which is a pseudovector with two
components. (Because the distinction between shear and reduced shear
is not important in the context of this paper, which is aimed at both
the astronomical and statistical communities, we refer to $g$ as
simply ``shear'' for convenience.)

Shear measurements are confounded
by several unavoidable observational effects.  First, for ground-based
telescopes, when the light passes through the atmosphere it is
convolved with a kernel that must be inferred from the data.  Second,
telescope optics (whether in space or on the ground) also cause the
image to be convolved with a kernel; this kernel may be more
predictable than the atmospheric kernel because the optics may be well
modeled.  In any case, the effective kernel imposed by atmosphere {\it
  and} optics is referred to as the point-spread function (PSF).
Third, emission from the sky causes a roughly
constant ``background'' level to be added to the whole image.  Fourth,
the detectors sum the light falling in each pixel, effectively
convolving the image with a square tophat window function,
and sampling the resulting image at the center of each pixel.  This
extra convolution
effect is treated by some authors as part of the PSF.  Fifth, the finite number of photons collected
in a given pixel is subject to Poisson noise (in addition the final detector readout adds Gaussian noise of zero mean, but this is ignored in GREAT08).

Thus a successful method must both filter the noise effectively and
remove the
significant PSF convolution kernel
in the observed galaxy
image.  To represent a method's ability to perform both tasks
in a single number for the GREAT08 Challenge,
we define a quality metric
\begin{equation}
Q = {10^{-4}\over \langle ( \langle g_{ij}^m - g_{ij}^t \rangle_{j\in k} )^2 \rangle_{ikl} }
\label{eq:Q}
\end{equation}
where $g_{ij}^m$ is the $i$th component of the measured shear for
simulation $j$, $g_{ij}^t$ is the corresponding true shear component,
the inner angle brackets denote an average over sets with similar
shear value and observing conditions $j \in k$, and the outer angle
brackets denote an average over simulations with different true shears
$k$, observing conditions $l$ and shear components $i$.

In our detailed discussion of the results below we also define a Q value for each simulation branch. In this case the average over different observing conditions $k$ is omitted
\begin{equation}
Q_l = {10^{-4}\over \langle ( \langle g_{ij}^m - g_{ij}^t \rangle_{j\in k} )^2 \rangle_{ik} }
\end{equation}
therefore
\begin{equation}
\frac{1}{Q} = \frac{1}{\langle Q_l \rangle_l} .
\end{equation}
This definition has the effect of strongly penalising methods that perform poorly in any single simulation branch, which is useful because the simulation branches are all chosen to be realistic scenarios in which we need to be able to measure good shears.
For a method to be used for all future analyses it must work
well on all branches of the simulations. In particular, there
are many small and low SNR galaxies that we would like to use
for cosmic shear cosmology.
However, the purpose of this results paper is to examine the performance of the different methods on the different branches in detail rather than relying on a single number $Q$ to differentiate between methods.

To set this metric in context, if a single constant value of zero
shear were submitted ($g^m_{1j} = g^m_{2j} = 0$ for all $j$) then
since
the rms true shear
$\sqrt{\langle g^{t2}_{ij} \rangle_{ij}} \sim 0.03$, Q would
have a value $\sim 0.1$.  To date, methods tested in STEP1
and STEP2 and used on real data have $Q \sim 10$
to $Q \sim 100$~\citep{kitchingmhvh08},
which is
sufficient for the surveys on which they were employed but
not
sufficient for
mid-term to far
future
surveys.

\cite{Amara:2007as} show that a
deep full-sky (e.g. Euclid-like)
 survey requires
that the additive error $c<0.0003$ and the multiplicative error
$m<0.001$. For a pure additive error this translates to a
requirement that $Q>1000$ and we set this as our target for
GREAT08 because additive errors are much more difficult to self-calibrate using pairs of tomographic redshift bins~\citep{Huterer:2005ez}~\citep[see also][]{VanWaerbeke:2006qt}.
A detailed analysis of the two separate terms is given in Appendix~\ref{appendix:Detailed_analysis_of_results}.

As defined, $Q$ penalises deviations from truth regardless of whether
they are random or systematic.  This is useful for selecting a winner,
but much can be learned by separating errors into random and
systematic parts.
For the systematic part we follow
STEP1 and STEP2
by defining
a multiplicative error $m$ and an additive error $c$
as the best-fit parameters to
\begin{equation}
 g^m_i - g^t_i = m_i g^t_i + c_i \,.
\end{equation}
We show some results for the average of the two components
$m=\langle m_i\rangle_i$,
$c=\langle c_i\rangle_i$
For a given method, changes in $m$ and $c$ across simulation branches
may indicate the strengths and weaknesses of the method.

Participants may optionally submit uncertainty estimates on their
shears.  These are compared to the
residuals of the
submitted shears over
sets of simulations with nearly identical true shear values.  If the
uncertainty estimates are wrong by more than a factor of two, the
submission is flagged as such, but is not penalised.  The main purpose
of GREAT08 is to produce a high Q value rather than yield correct
uncertainty estimates.

A method is not useful if it obtains very small shear biases at the expense of throwing away most of the information and thus very noisy shear estimates. The quality factor $Q$ will be worse if a method has very noisy shear estimates because the rms difference between the truth and submission will be non-negligible even if the biases are zero. We therefore calculate the scatter of the submitted shear values about the best linear fit to the true shears. Specifically, we plot submitted $g_1$ values as a function of true $g_1$, with one point for each {\tt FITS} file and fit the straight line described above. We find the rms residual to obtain the scatter $\sigma_1$ in the first component $g_1$. We repeat for $g_2$ and write $\sigma\equiv\langle \sigma_i \rangle_i$ averaging over the two shear components $i$.
See~\cite{kitchingmhvh08} for additional discussion.

\section[]{Methods}
\label{sect:Methods}

\begin{table*}
\begin{center}
\begin{tabular}{l@{}ccccc}
\hline\hline
\bf{Participant(s)}        & \bf{Key} &
\bf{Action 1} & \bf{Action 2} & \bf{Action 3} \\
\hline\hline
Hosseini, Bethge & HB &
Estimate power spectrum & Average power spectra & Fit elliptical model $*$ PSF
\\ \hline
Lewis & AL &
Estimate centroids & Average images & Fit elliptical model $*$ PSF
\\ \hline
Kitching & TK$^\dagger$ &
Fit elliptical model $*$ PSF &
Combine ellipticity PDFs &
Calculate shear
\\ \hline
Heymans & CH$^\dagger$ &
Measure weighted quadrupole moments & Correct for weight and PSF & Average shear estimates
\\ \hline
Paulin, Gentile & PG &
Fit elliptical model $*$ PSF & & Average shear estimates
\\ \hline
Velander & MV &
Fit flexed elliptical model $*$ PSF & & Average shear estimates
\\ \hline
Kuijken            & KK$^\dagger$       &
Fit elliptical model $*$ PSF && Average shear estimates
\\\hline
Harmeling, Hirsch, Sch\"{o}lkopf & HHS3 &
Estimate centroids & Average good images & Fit elliptical model * PSF
 \\ \hline
Bridle & SB$^\dagger$ &
Fit elliptical model $*$ PSF & & Average shear estimates
\\ \hline
Harmeling, Hirsch, Sch\"{o}lkopf & HHS2 &
Estimate centroids & Average images & Fit elliptical model $*$ PSF
\\ \hline
Harmeling, Hirsch, Sch\"{o}lkopf & HHS1 &
Fit elliptical Gaussian & Correct for model and PSF & Average shear estimates
\\ \hline
Jarvis             & MJ$^\dagger$    &
Fit ``elliptical" model $*$ PSF & & Average shear estimates
\\ \hline
Bridle, Schrabback & USQM$^\dagger$ &
Measure quadrupole moments - PSF & Average quadrupole moments & Calculate shear \\ \hline
\hline
\end{tabular}
\end{center}
\caption{Table of participants, figure legend identifiers and pseudo-code which attempts to summarise the main actions carried out in each method. ``$*$ PSF" indicates that a PSF convolved model was fitted. ``PDF" stands for probability density function. Daggers after the Key indicate GREAT08 Team entries. More information is provided in the main text and in Appendix~\ref{appendix:Additional_information_on_methods}.
\label{tab:participants}}
\end{table*}

In this section we briefly summarise the algorithm used by each
submitting group. Table~\ref{tab:participants} lists the participants, their methods, and the corresponding identifiers used in subsequent tables and in the figure legends. Methods with an asterisk indicate GREAT08 Team entries; these participants had access to the internal details of the GREAT08 Challenge simulations, but they did not consciously use this information in their analyses. Entries from PG, MV had some overlap with the GREAT08 Team. Not all submitting groups submitted results for both types of Blind simulation.
An additional table (Table~\ref{tab:participants_more}) in Appendix~\ref{appendix:Additional_information_on_methods} gives further information including urls where more information can be found.

\begin{figure}
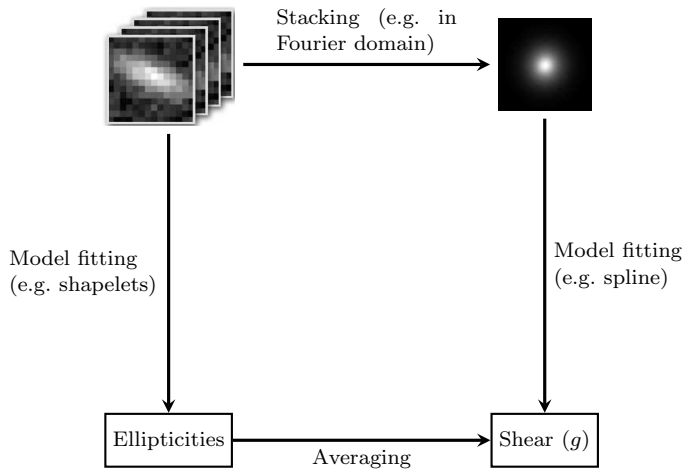

\center
\include{great08figure}
\caption{
Illustration of the different routes to a combined shear statistic from multiple galaxies. The lower left route is the traditional approach in which each galaxy image is analysed separately to produce a shear estimate. The upper right route illustrates the ``stacking" methods which average some statistic of each image and perform shear estimation on the averaged statistic.}
\label{fig:stacking}
\end{figure}

For a quick overview we attempt to summarise each method with just three
action steps in Table~\ref{tab:participants}.
We see that a key differentiating factor is the stage at which an average is performed over galaxies in the image.
HB, AL and USQM
as ``stacking" methods hereafter. The two different routes are illustrated in Fig.~\ref{fig:stacking}.

STEP2 classified methods according to their methods for PSF correction and construction of a shear estimator. PSF ``deconvolution'' methods convolve a model with the PSF before fitting as indicated by ``$*$ PSF'' in the table; PSF ``subtraction'' methods subtract a contribution due to the size and ellipticity of the PSF. ``Active'' shear measurement methods sheared a ``circular'' galaxy model until it best matched the data, generally indicated by the word ``fit" in the action list; ``passive'' methods constructed a shear estimator from a combination of shape statistics and an estimate of how these would further change under a shear. This classification system proved insufficient to capture the more varied behaviour of methods containing new ideas in GREAT08.
We next summarise each method in turn, in order of decreasing $Q$ value
on RealNoise\_Blind.

{\it HB:}
The magnitude of the Fourier transform of the galaxy image raised to an arbitrary power is a characteristic feature of the individual galaxies. This feature is independent of the spatial location of the galaxy center to a high precision, provided that the smoothed galaxy intensity decays sufficiently fast
towards the edge of the image. No other assumptions are necessary. Because the galaxy images are contaminated by Poisson noise, an unbiased estimator of the power spectrum is given by the power spectrum of the noisy image minus a constant. The resulting image obtained by averaging over the unbiased estimators of the individual galaxy power spectra is an elliptically contoured function multiplied by the power spectrum of the convolution kernel plus Gaussian noise. After suitable normalization, the square root of the covariance matrix of the elliptically contoured function is equal to the shear coordinate transformation matrix. For parameter fitting, HB used a weighted non-linear least square method for which the weights are equal to the inverse of the standard deviation of the noise. For more information see \cite{hosseinib09}.

{\it AL:} This method was inspired by \cite{Kuijken99} and is described in \cite{Lewis09}. Centroids for each galaxy are determined and all galaxies in a {\tt FITS} image are stacked on a sub-pixel scale. A
PSF convolved
elliptical profile is fitted to this stacked image, and the ellipticity corresponds to the shear. As pointed out in \cite{Lewis09}, the advantage of this approach is that the individual non-elliptical shapes of individual galaxies are averaged out. This fact was taken advantage of in HB, HHS2 and HHS3.

{\it TK:} The {\it Lensfit} code
fits a sum of co-elliptical
exponential and de Vaucouleurs models to each individual galaxy and
the best fit ellipticity is found.
The bulge (de Vaucouleurs component) to disk (exponential component) fraction is a free parameter in the fit.
The shear is calculated using a Bayesian estimator. For more details see Appendix F of the GREAT08 Handbook and also \cite{millerkhhv07} and \cite{kitchingmhvh08}
The version used here differs from the previously published implementations by including subpixel estimation of galaxy positions and adaptive ellipticity grid refinement.

{\it CH:} An implementation of the longstanding KSB \citep{KSB}
method, which is the most widely used code on observational data.
For more information, see Appendix C of the GREAT08 Handbook.

{\it PG:} 
For each galaxy, a 6-parameter Sersic model is convolved with the PSF and pixellated. This is fitted to the image through $\chi^2$
minimization using the gradient-expansion algorithm by LevenbergMarquardt. The six fitted parameters are: the centroid (2 parameters),
the magnitude, the size, and the ellipticity (2 parameters). The estimated shear of an individual galaxy is derived from its fitted
parameters and the averaged shear over a number of galaxies is the average of individual shears.

{\it MV:}
This method is an extension of the KK method described below. It is being developed with the aim of measuring higher order galaxy image distortions, known as flexion, as well as shear. These higher order distortions add important detail to the measurement of galaxy halo density profiles and to dark matter mapping. For more information on this method see Velander \& Kuijken in prep. and for further detail on flexion see \cite{bacongrt06}.

{\it KK:}
Each individual galaxy is modelled as a sheared, circular source described by means of the first-order shear operators in shapelet space. The PSF is also modelled as a high-order shapelet expansion, and all convolutions are carried out in shapelet space using the prescriptions in~\cite{refregier03_shapeletsI}.  For further information see \cite{kuijken06} and Appendix D of the GREAT08 Handbook.

{\it HHS1/HHS2/HHS3:}
In {\it HHS1} an elliptical
Gaussian is
fitted to each galaxy image by minimizing the mean-squared error via
gradient descent in the 6 model parameters.
As in {\it SB}, the average ellipticity is
taken as an estimate for the shear. Due to the simplified galaxy
model and the PSF blur a systematic bias is introduced, which is corrected for
by off-setting the
ellipticity values
and via
calibration using the training data.
The methods {\it HHS2} and {\it HHS3} aim to
be more robust by adopting the idea of {\it AL} to stack all galaxy images within one {\tt FITS} file on a
subpixel scale in order to increase the SNR. In addition, in {\it
HHS3} corrupted images were
removed before
stacking.

{\it SB:} The {\it im2shape} code models each individual galaxy as a sum of co-elliptical Gaussians. The parameters are marginalised using MCMC sampling and the mean ellipticity of the samples is taken to correspond to the shear.
For computational speed, only 16$\times$16 pixels in the center of each postage stamp were used in the fit. See Appendix E of the
GREAT08 Handbook and \cite{bridlekbg02}.

{\it MJ}: This algorithm seeks a coordinate system in which a model
of the galaxy is found to be round. The model is convolved by
the PSF and then compared to the observed pixel intensities. A
shapelet decomposition is used for the underlying model, and
roundness is defined as the second order shapelet coefficients
being 0. Then the shear that brings this coordinate system back
to the actual observation is assigned as the shape of the
galaxy. For more information see \cite{bernsteinj02}, \cite{nakajimab07} and Appendix D of the GREAT08 Handbook.

{\it USQM:} This is a very simple method, not actually used in practice, but provided as a baseline comparison. The unweighted quadrupole moments of each galaxy are calculated within a square aperture of 20 pixels by 20 pixels. These are averaged (stacked) over all galaxies in each {\tt FITS} image and the PSF is removed by subtracting the PSF quadrupole moments. See Appendix B of the GREAT08 Handbook for more information.

In terms of the nomenclature introduced in STEP2 most of the methods forward fit an elliptical PSF convolved model (``active", ``deconvolution").
This is in contrast to the situation in STEP1 and STEP2 where
the majority of the methods were ``passive" PSF subtraction methods.
There were no stacking methods in STEP1 or STEP2.

\section[]{Results}
\label{sect:Results}

There were two blind challenges: LowNoise$\_$Blind contains high
SNR images and RealNoise$\_$Blind contains images with a realistic noise level.  The GREAT08 Challenge prize for highest $Q$ value is
based on the RealNoise$\_$Blind results.
The LowNoise$\_$Blind competition contained significantly less data and should have been an easier challenge. Furthermore, the galaxy properties in LowNoise$\_$Blind were similar to those in RealNoise$\_$Blind and are mostly co-centered bulge plus disk models. It could therefore have been useful to optimise some properties of methods on the LowNoise$\_$Blind images in preparation for RealNoise$\_$Blind.
First, we examine the
LowNoise$\_$Blind results.

\subsection[]{LowNoise Blind Results}
\label{sect:LNB_results}

\begin{figure}
\center
\epsfig{file=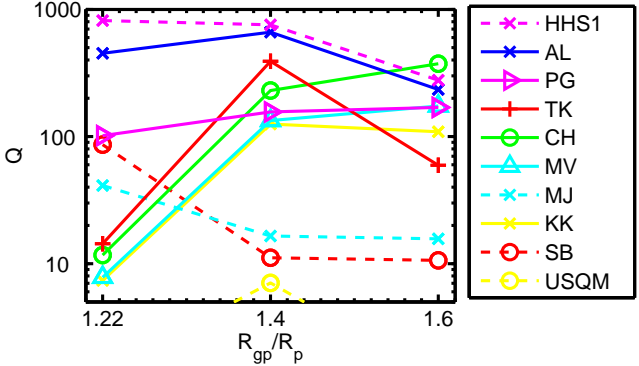,height=5cm,angle=0}
\caption{
Our figure of merit
Q as a function of galaxy size
for LowNoise$\_$Blind.
}
\label{fig:Q_breakdowns_LNB}
\end{figure}

\begin{table}
\begin{center}
\begin{tabular}{l@{}ccl}
\bf{Rank} & \bf{ID}  & \bf{Method} &
\bf{Q}  \\\hline\hline
1 &  HHS1 & Gauss &
488 \\ \hline
 2 &     AL &    CLT KK99 &
375 \\ \hline
 3 &     PG & gfit &
 136\\ \hline
 4 &     TK &   Lensfit &
     33.7 \\ \hline
 5 &     CH &   KSBf90 &
     32.4 \\ \hline
 6 &     MV &    KKshapelets with flexion &
    21.2 \\ \hline
 7 &     MJ &   BJ02 deconvolved shapelets&
   20.2 \\ \hline
 8 &     KK &   KKshapelets &
 19.7 \\ \hline
 9 &     SB &   im2shape &
  15.3\\ \hline
 10 & USQM &   USQM &
 1.84 \\ \hline\hline
\end{tabular}
\end{center}
\caption{LowNoise$\_$Blind leaderboard at the close of the challenge.
See Table~\ref{tab:participants} and Section~\ref{sect:Methods} for more information about each method.
  \label{tab:LNBleaderboard}}
\end{table}

Table~\ref{tab:LNBleaderboard} shows the LowNoise$\_$Blind leaderboard
at the close of the challenge.  The winner in LowNoise\_Blind is the
Gauss method of S. Harmeling, M. Hirsch, and B. Sch\"{o}lkopf.
The top three methods in LowNoise$\_$Blind are
not
GREAT08 Team methods.
Note that HB did not submit a result for LowNoise$\_$Blind.

Fig.~\ref{fig:Q_breakdowns_LNB} shows our shear measurement figure of merit $Q$ as a function of
the ratio between the convolved galaxy size and the PSF size,
 ${R_{gp}/  R_p}$.
Since the number of galaxies decreases steeply as a function of galaxy size in real data, it is desirable to have a shear measurement method that allows the use of small galaxies.
It is often assumed that shear measurement biases are larger for small galaxies. There are some examples where this is true in STEP2
Fig. 7, and~\cite{nakajimab07} Fig. 5. However the shear biases are caused by a combination of two effects: a poorly measured PSF and inherent biases that exist even if the PSF is perfectly known.
It is expected that an incorrect PSF model will affect small galaxies the most, since for the largest galaxies the PSF has little effect \citep[e.g. Eq. 13 of][]{paulinavrb08}.
In GREAT08 the exact PSF equation is known and if this information is properly used then the results will tell us about the inherent biases, for which there are less clear expectations.

HHS1 (dashed magenta line in Fig.~\ref{fig:Q_breakdowns_LNB}) is
the clear winner overall in LowNoise\_Blind and wins at both the fiducial and small galaxy sizes. The implementation of KSB
by CH (solid green line in Fig.~\ref{fig:Q_breakdowns_LNB}) provided
the best performance for highly resolved galaxies.
As discussed above, this general trend of increasing $Q$ with increasing galaxy size was expected, and is followed for many methods.
The winning method HHS1 performed worse as the galaxy size increased for LowNoise\_Blind.
We suggest that
the method for calibrating the ellipticities for the PSF blurring was less reliable at large galaxy sizes due to the fact that the large elliptical galaxies sometimes extend beyond the $39\times39$ pixel postage stamp.

Further analysis of the LowNoise$\_$Blind results in terms of multiplicative and additive shear calibration biases can be found in Appendix~\ref{appendix:Detailed_analysis_of_results:LNB}.

\subsection[]{RealNoise Blind Results}
\label{sect:RNB_results}

\begin{figure*}
\center
\epsfig{file=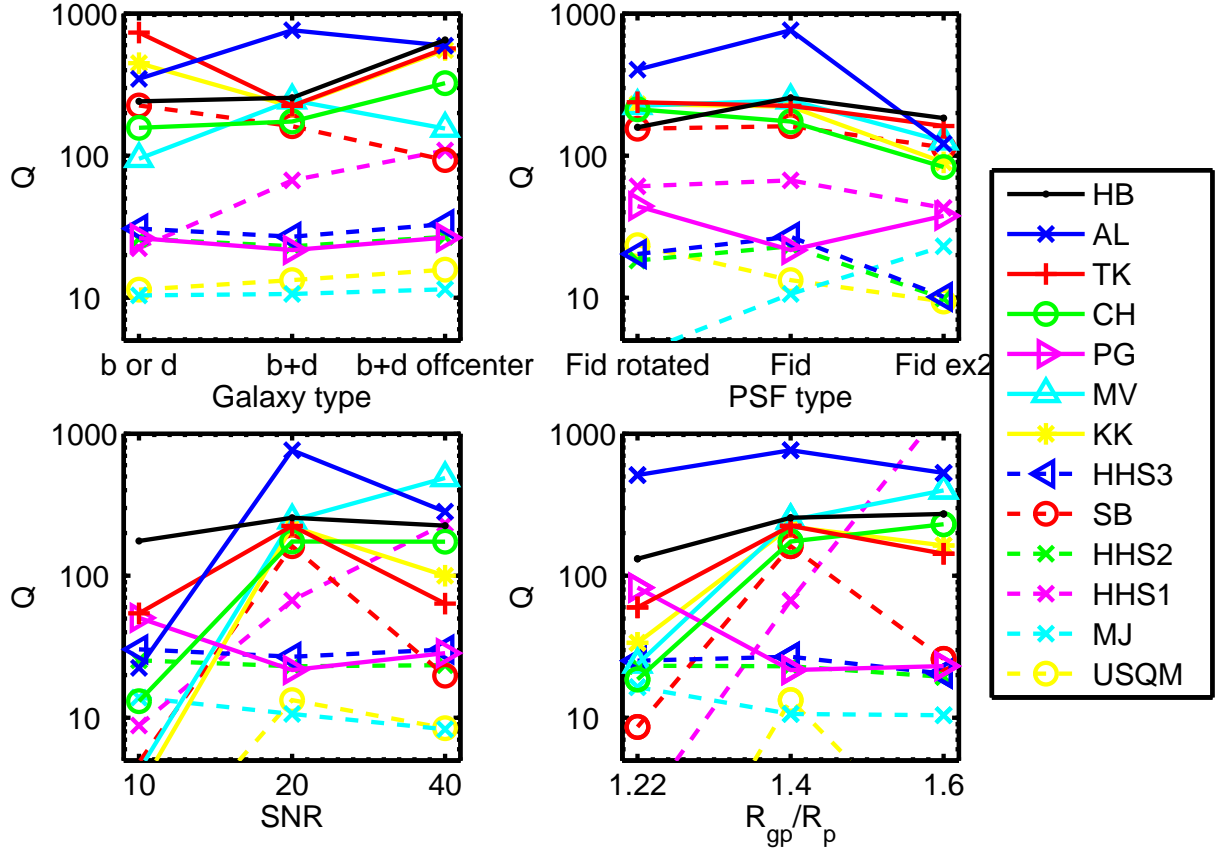,height=11.5cm,angle=0}
\caption{
Shear measurement figure of merit
$Q$ as a function of simulation properties for RealNoise\_Blind.
}
\label{fig:Q_breakdowns_RNB}
\end{figure*}

\begin{table}
\begin{center}
\begin{tabular}{l@{}ccl}
\bf{Rank} & \bf{Author}  & \bf{Method} &
\bf{Q}  \\\hline\hline
1 &     HB &    CVN Fourier &
 211 \\\hline
2 &     AL &    KK99 &
 131     \\\hline
3 &     TK &    Lensfit &
  119 \\\hline
4 &     CH &   KSBf90 &
  52.3    \\\hline
5 &     PG &    gfit&
 32.0 \\\hline
6 &     MV &    KKshapelets with flexion&
    28.6 \\\hline
7 &     KK &   KKshapelets &
 23.0 \\\hline
8 &     HHS3 &  GaussStackForwardGaussCleaned &
22.4\\\hline
9 &     SB &   im2shape &
 20.1 \\\hline
10 &    HHS2 & GaussStackForwardGauss&
19.9\\\hline
11 &    HHS1 & Gauss &
12.8 \\\hline
12 &    MJ &   BJ02 deconvolved shapelets &
 9.80 \\ \hline
13 &    USQM &        USQM &
1.22 \\ \hline\hline
\end{tabular}
\end{center}
\caption{RealNoise$\_$Blind leaderboard at the close of the challenge.
  \label{tab:RNBleaderboard}}
\end{table}

The main challenge consisted of 27 million galaxies with roughly a factor of 10 more noise per pixel, corresponding to the type of image that we will ultimately want to use for cosmic shear.
The RealNoise$\_$Blind leaderboard at the close of the challenge is shown in
Table~\ref{tab:RNBleaderboard}.  The winner of the GREAT08
Challenge is clearly the `CVN Fourier' method by R. Hosseini and M. Bethge, HB.
This method was inspired by the second-place AL method, but improves
on a key limitation which was highlighted by \cite{Lewis09}
in that it did not depend on the galaxy centroid.

Fig.~\ref{fig:Q_breakdowns_RNB} shows $Q$ as a function of galaxy type,
PSF type, SNR, and
galaxy size
for
RealNoise\_Blind.
The central, fiducial, value is the same on each of the four panels.
Each point on the panels corresponds to a single set of conditions; for example, for the SNR$=10$ point, all other parameters are set at the fiducial value.

HB performs consistently well through all branches of the simulation, with significantly improved
performance on the ``b+d offcenter'' galaxies.
AL actually outperformed HB on six
of the nine simulation branches, and obtains a $Q$ value a factor of almost 4 larger than any other method for the fiducial simulation set,
which is
close to our target value of 1000.
AL was second overall mostly as a
result of a poor performance on the low SNR branch,
and to a lesser extent on the
``Fid $e \times$ 2'' PSF.
It would be interesting to see if the results could be improved in either of these regimes, for example with better centroiding at low SNR or better modeling of the ``Fid $e \times$ 2'' PSF.

TK uses a model with coaligned exponential and de Vaucouleurs components which explains why the results on `b or d' are so good.
It also does well on `b+d offcenter'. If the galaxy model could be extended then this may improve the other results, which all use the fiducial galaxy type.
KK also performs well on the ``b or d'' branch, and to a lesser extent, so does SB. Both these methods also assume galaxies have elliptical isophotes, which matches exactly the model in the simulation.

The best method at the high SNR end of RealNoise\_Blind is MV (KK shapelets with flexion), which also performs well for the larger galaxies.
HHS1 on the larger galaxy
branch is the only method on any branch to achieve greater than the $Q\sim 1000$
level required for future precision surveys.
This trend is surprising given that it reverses the trend with ${R_{gp}/ R_p}$ seen in LowNoise\_Blind.
It also obtains a good $Q$ value at the high SNR end (SNR=$40$) of RealNoise\_Blind, which is
not surprising given the
strong
performance in LowNoise\_Blind (SNR$=200$).

Note that the absolute value of $Q$ will depend on the noise on the shear measurements and on the number of realisations over which the average is performed. Therefore it is not terribly meaningful to compare values between LowNoise and RealNoise, however the $m$ and $c$ values can be usefully compared. These values are discussed for RealNoise\_Blind in Appendix~\ref{appendix:Detailed_analysis_of_results:RNB}.

\section{Discussion}
\label{sect:discussion}

The GREAT08 Challenge has moved shear measurement research significantly beyond STEP1 and STEP2.
We recognised that the shear measurement problem is intrinsically a statistical, not astronomical, problem and wrote a description addressed at non-astronomers (the GREAT08 Handbook).
At the launch of the challenge we had achieved the following:
\\$\bullet$ We moved from end-to-end simulations to simpler simulations which isolate a key difficult part of the shear measurement problem without confusion from other effects.
\\$\bullet$ The simulations focus in on key areas of simulation parameter space and allow a detailed assessment of the success of different methods in the various regimes explored.
\\$\bullet$  We used a larger suite of simulations to assess methods at a much higher level of precision than was possible in STEP1 and STEP2; this level of precision is appropriate for
the most ambitious planned cosmic shear surveys.
\\$\bullet$
The GREAT08 Team was formulated from the original STEP Team and new
groups e.g. LensFit were incorporated and assessed as part of the blind competition.
\\$\bullet$ We formulated a new figure of merit with which to assess the results of the challenge and provided active leaderboards during the challenge.
\\$\bullet$ The GREAT08 Team codes were all made publically available at the launch of the challenge.

In addition to the six GREAT08 Team entries on the leaderboards at the start of the challenge there were five new entries which included
computer scientists and non-lensers. The GREAT08 Challenge has therefore achieved its main goal of reaching out beyond the existing shear measurement community.

The GREAT08 Challenge prize for the highest $Q$ value in RealNoise\_Blind went to Reshad Hosseini and Matthias Bethge (HB).
The GREAT08 Team
also awarded a prize for a significant contribution to advancing shear
measurement methods to Antony Lewis (AL), specifically for superb
results over a significant range of simulation branches, and a
timely summary of the problem that highlighted important issues
\citep{Lewis09}.
Neither of these prizewinning groups are associated with existing lensing groups.

The shear measurement problem has been invigorated by the Challenge and by the new ideas brought in. The most important new ideas are
\\$\bullet$ a consideration of the impact of the assumed galaxy model on the accuracy of shear measurements;
\\$\bullet$ a reconsideration of the stage in the measurement process at which to average observational quantities.

The assumed galaxy model has recently been shown to be important in causing biases in shear measurement~\citep{Lewis09,voigtb09,melchiorblb09}. The existence of this bias was first pointed out by~\cite{Lewis09} and this was the motivation for using a ``stacking" method by both AL, HB and HHS2/3. In both methods the individual galaxy properties are averaged away before a model is fitted, by averaging together simple statistics of the galaxy images. AL pointed out that averaging together the images themselves is not fully independent of the galaxy model, the PSF or the shear because a centroid must be estimated before stacking. HB solved this by instead stacking two-point statistics of the image (specifically the power spectrum), which is insensitive to the centroid. This raises the general question of what quantity should be averaged (or otherwise combined), and at what stage, when presented with many galaxy images all with the same shear value.

The success of the stacking methods
on images with constant galaxy properties leads to questions about
how well stacking could work on more realistic data.
Because shear varies with position in real data, the stacking process
will
average
the shear signal as well as
nullify
the observation effects
it was designed to remove.
However,
we speculate that
the average shear in a patch of sky is still a useful cosmological quantity, as has sometimes been considered
\citep[e.g. most recently the top hat shear variance statistic shown in Fig. 5 of][]{fuea08}
\citep[see also cosmic shear ring statistics described in ][]{schneiderk07,eiflersk09}.
For lensing analyses of clusters or galaxies,
the assumption of axisymmetry is often made
which
lends itself
naturally to stacking
in annuli about the center of the cluster.
It would also be necessary to determine
how to properly stack
galaxies with a range of SNR
or PSF
in a given patch of sky, and especially
how to tackle galaxies with a range of redshifts, and thus a range of
shears.
For example, 3D lensing~\citep{heavens03,kitchinghvsm08} is specifically designed to take into account the probability distributions in redshift and shear for each galaxy separately.

The results of GREAT08 show that different methods are successful in different corners of parameter space and many results are close to the target $Q$ value of 1000.
The results from different simulation branches give clues as to where methods could be improved and we expect to see further work on developing the methods. The winning method HB only finished its first run two days before the challenge deadline and therefore
it could be optimised further. In addition it shows remarkably stable performance as a function of SNR implying that the good $Q$ results might continue down to even lower SNR values. On the fiducial simulations AL achieved a $Q$ value nearly four times higher than previous work, marking a significant improvement. The performance at low SNR is the clear next area for investigation for this method.
TK obtains good results, in particular when the underlying model was
similar to the model in the simulation.

GREAT08 marks the first in a series of GREAT challenges,
which are intended to be
a roadmap of simulations leading
up to the
real
grand observational challenges that the community will face with the next generation of cosmic shear surveys.
The next challenge
in the series
will be GREAT10. This will represent the next step towards creating fully realistic simulations.
Many aspects of the GREAT10 simulation will be familiar
from
GREAT08, though they
will differ in some key aspects. The most significant change will be spatial variation: both the shear and
PSF
will vary across each image. GREAT10 will also invite people to solve an extra
cosmic shear challenge,
estimating the convolution kernel from images to sufficient accuracy. For more information on GREAT10
visit http://www.great10challenge.info.

\section*{Acknowledgments}
We thank the PASCAL Network for support.
We thank the GREAT08 Team and participants at the GREAT08 Mid-Challenge Workshop and GREAT08 Final Workshop including
Hakon Dahle, Domenico Marinucci and
Uros Seljak.
We thank the organisers of Cosmostats09 for hosting the GREAT08 Challenge Final Workshop within Cosmostats09 in Ascona.
We thank the Aspen Center for Physics where part of this work was carried out.
We are grateful to Jeremy Yates for help with setting up the GREAT08 server.
SLB thanks the Royal Society for support in the form of a University
Research Fellowship.
TDK is supported by STFC Rolling grant RA0888.
JR is supported in part by the Jet Propulsion Laboratory, which is run by Caltech under
a contract from NASA.
MS was supported in part by the program \#11288 provided by NASA through a grant from the STScI, which is operated by the Association of Universities for Research in Astronomy, Inc., under NASA contract NAS 5-26555.

\bsp

\bibliographystyle{mn2e_eprint}

\appendix

\section{Details of the image simulations}
\label{appendix:Details_of_image_simulations}

\subsection{PSF models}
\label{appendix:PSF_models}

\begin{table}
\caption{
PSF ellipticities.
}
\label{tab:psf_e_values}
\center{
\begin{tabular}{|l|c|c|c}
\hline
PSF type        & Filename & $e_1$ & $e_2$ \\
\hline
Fid & set0001 & -0.019 & -0.007 \\
Fid rotated & set0002 & 0.007 & -0.019\\
Fid $ e \times 2$ & set0003 & -0.038 & -0.014 \\
\hline
\end{tabular}
}
\end{table}

In an attempt to isolate problems in the shear estimation pipelines and make the challenge more accessible
we provided maximal information about the PSFs used during the competition.

The PSFs had a truncated Moffat profile
\begin{equation}
I_{p}(r)
=
\left\{
\begin{array}{ll}
  \left(1 + \left(\frac{r}{r_d}\right)^2 \right)^{-\beta}
  \hspace{1cm}&
   \mbox{ $r<r_c$}
  \\
 0 & \mbox{ $r>=r_c$}
\end{array}
\right.
\label{eq:moffat}
\end{equation}
where we set $\beta=3.5$.
This profile is motivated by the combination of diffraction limited
optics with random Gaussian blurring by the atmosphere and is therefore
reasonably representative of PSFs for ground-based telescopes.
The scale radius $r_d$ was determined by setting the Full Width at Half Maximum (FWHM) to 2.85 pixels. $r_c$ was set to twice the FWHM.
Three different PSFs were used in the GREAT08 Challenge, each with a different ellipticity, as shown in Table~\ref{tab:psf_e_values}.

Star catalogues consisted simply of the position of the point source.
The x positions were drawn from a Gaussian of standard deviation 1.2 pixels centered on the middle of the postage stamp, similarly for the y positions.
The star catalogues were provided at the time of the challenge.
The convolution kernel and image generation method are described below.

\subsection{Galaxy catalogue generation}
\label{appendix:Galaxy_catalogue_generation}

The information provided in this appendix subsection was not available during the Challenge.

In general, the galaxies in GREAT08 are the sum of two components, each with a Sersic~\citep{1968adga.book.....S} intensity profile
\begin{equation}
I(r) =
\left\{
\begin{array}{ll}
 I_o \exp\left(-\kappa (r/r_e)^{1/n} \right) \hspace{0.5cm}
 & r<4r_e
 \\
 0 &  r>=4 r_e
\end{array}
\right.
\label{eq:sersic_law}
\end{equation}
where $I(r)$ is the amount of light per unit area at a radius $r$, and $\kappa \simeq 2n-0.331$ \citep[see e.g.][]{penghir02}. The scale radius $r_e$ and the total intensity (which determines $I_o$) are free parameters specified in the catalogues.
The first component, with $n=4$, is an approximation to the central bulge component of galaxies, corresponding to a de Vaucouleurs profile.
The second component, with $n=1$, is an approximation to the exponential disk component of galaxies.
Circular galaxy images are made according to the profile $I(r)$ described above and then distorted according to the galaxy ellipticity and shear as described below.

The x and y positions of the bulge component were each drawn from a
Gaussian of standard deviation 1.2 pixels centered on the middle of the postage
stamp. By default the positions of the disk component were set equal
to those of the bulge, except in one branch of the
RealNoise\_Blind simulations, as described below (see
Table~\ref{tab:RealNoise_Blind_pars}).

For each object, the total flux (integral of $I(r)$ over the postage stamp)
in the disk component, as a fraction of the total flux in both
components, is in general a random number drawn from a uniform
distribution between 0 and 1.  However, for LowNoise\_Known,
RealNoise\_Known, and one branch of RealNoise\_Blind, this fraction was
set to {\it either} 0 {\it or} 1. So, in these simulations, the galaxies had  either a pure de Vaucouleurs or pure exponential profile.

\begin{table}
\caption{
Galaxy scale radius values for single-component galaxy models.
The left hand column gives the ratio of PSF convolved galaxy FWHM to the PSF FWHM. The middle column gives the scale radius for a single component disk model. The right hand column gives the scale radius for a single component bulge model. These values are interpolated to produce scale radius values for two-component models, as described in the text.
}
\label{tab:re_values}
\center{
\begin{tabular}{|c|c|c|}
\hline
$R_{gp}/R_p$
& Disk $r_{e,d0}$ & Bulge $r_{e,b0}$\\
\hline
1.22 & 0.82 & 1.59\\
1.4     & 1.3 & 3.8\\
1.6     & 2.4 & 18.0\\
\hline
\end{tabular}
}
\end{table}

The scale radii $r_e$ of each component were set by considering high
resolution circular galaxy images after convolution with the
appropriate PSF.  For single-component models (i.e. when the bulge to
total flux is zero or unity), $r_e$ is set such that the convolved
image has a FWHM of 1.4 times that of the PSF, $F_{gp}=1.4F_p$, in the
fiducial branch.  Values 1.22 or 1.6 were used for some other branches
to explore the effect of galaxy size, as detailed below (see Tables
~\ref{tab:LowNoise_Known_pars} and~\ref{tab:RealNoise_Blind_pars}).
The resulting $r_e$ values for single-component models are provided in
Table~\ref{tab:re_values}.  For two-component models the disk scale
radius is a set multiple of the bulge scale radius, $r_{e,d}=2 r_{e,b}
* r_{e,d0}/r_{e,b0}$ using values from Table~\ref{tab:re_values}.  The
bulge scale radius was set by simulating a high resolution
two-component circular model with the required bulge to total flux
ratio and finding the value such that the FWHM had the required value
(by default 1.4 times the PSF FWHM).

The ellipticities of the bulge and disk were drawn from
\begin{eqnarray}
P(\epsilon)&=&\epsilon\left(\cos\left(\frac{\pi\epsilon}{2}\right)\right)^2
\exp \left(-\left(\frac{2\epsilon}{B}\right)^C \right)
\label{eq:P_of_epsilon}
\end{eqnarray}
with $B=0.05$, $C=0.58$ for the bulge and $B=0.19$, $C=0.58$ for the disk;
$\epsilon\equiv(a^2-b^2)/(a^2+b^2)$ where $a$ and $b$ are the major and minor axes respectively.
Since ellipticities close to unity become unphysical, we truncate the distribution at $\epsilon=0.9$ and set all objects with $\epsilon>0.9$ to have $\epsilon=0.9$.
This distribution was loosely motivated by results from the APM survey
\citep{crittendennpt01};
The bulge and disk ellipticities are drawn independently from the above distributions and are thus uncorrelated.
The angle between the bulge major axis and the positive $x$ axis is drawn from a uniform distribution between 0 and $180$ degrees. The disk angle is equal to the bulge angle but perturbed by a Gaussian of standard deviation 20 degrees.

Five thousand galaxy parameters were simulated per image set by
drawing from the above distributions. To minimise noise the parameters
were all rotated by 90 degrees to produce the remaining 5000 galaxy
parameters. (i.e. all angles are increased by 90 degrees, $x$
positions become $y$ positions, and $y$ positions become negative $x$
positions.)  The list was randomised to hide the pairings.  This
paired rotation was introduced in STEP2 to reduce shape noise.  In the
absence of a PSF or shear the shear estimates from each galaxy in a
pair are expected to cancel, thus removing noise arising from the
intrinsic ellipticities of galaxies.

Signal-to-Noise Ratios (SNR) are assigned in the catalogues and are used during image simulation to set the flux in the galaxy image. For LowNoise images the value is 200, and for RealNoise images the default value is 20, with variations to 10 and 40 within RealNoise$\_$Blind. The definition of this number in terms of the noise model is described in the following subsection.

For LowNoise$\_$Known and RealNoise$\_$Known the galaxies all have just a single component and within each set, each galaxy is assigned a de Vaucouleurs or an exponential profile at random.
The galaxies in LowNoise$\_$Blind all have a bulge plus disk two-component model as described in the text above. The majority of the galaxies in
RealNoise$\_$Blind have the same two-component model as in LowNoise$\_$Blind.
One of the nine RealNoise$\_$Blind branches has single-component galaxies as in the Known simulations.
The two-component models all share the same centroid for the bulge and disk, except for one of the nine RealNoise$\_$Blind branches, in which the bulge is off-centered from the disk by a Gaussian of standard deviation 0.3 pixels.

\begin{figure*}
\center
\epsfig{file=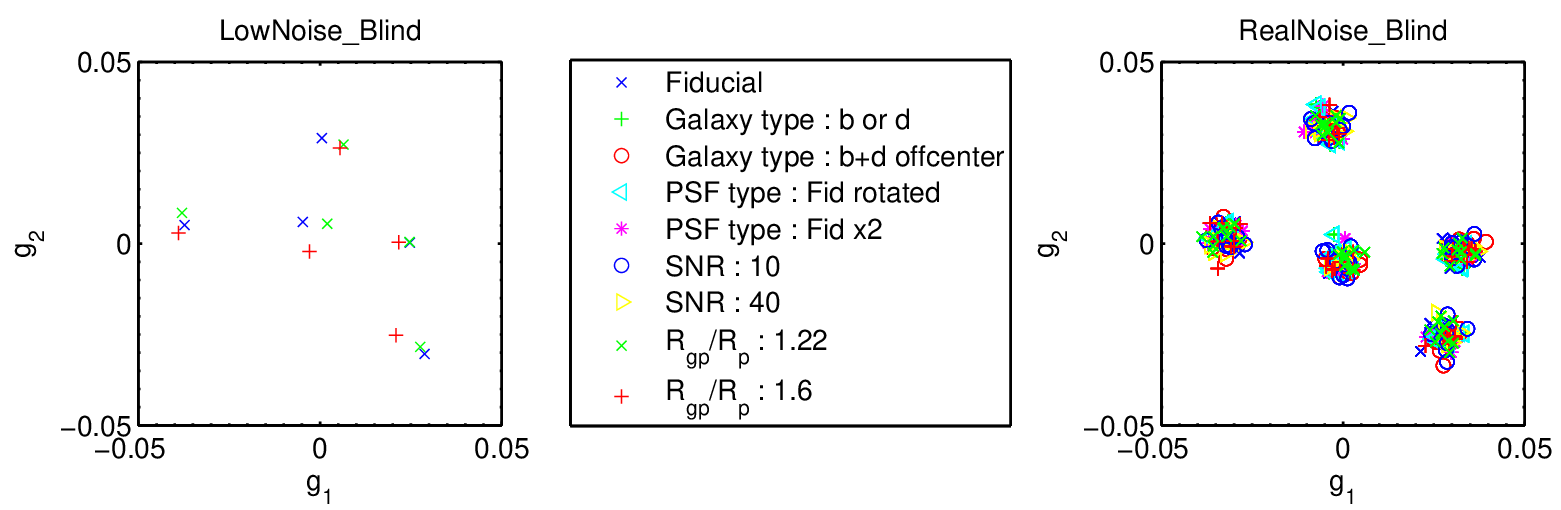,width=18cm,angle=0}
\caption{
True shears for LowNoise$\_$Blind and RealNoise$\_$Blind, color coded for the different branches of the simulations. }
\label{fig:g1g2_truth}
\end{figure*}

The true shears for LowNoise$\_$Known and RealNoise$\_$Known
were provided throughout the challenge. They are Gaussian distributed with a standard deviation of 0.03 in each of $g_1$
and $g_2$, and zero mean.
The true shears for LowNoise$\_$Blind and RealNoise$\_$Blind have now been released, and are illustrated in Fig.~\ref{fig:g1g2_truth}. These shears are perturbations around the root values
$g_1=(-1, 0, 1, 0, -1/\sqrt{2})\times 0.037$ and
$g_2=(0, 0, 0, 1, -1/\sqrt{2})\times 0.037$
and thus do not have zero mean. This distribution is chosen instead of a Gaussian to improve the uncertainties on linear fits to the output versus true shear.
For LowNoise$\_$Blind, one position in shear space is drawn from around each root and there is one set with this shear.
For RealNoise$\_$Blind, 50 positions in shear space are drawn from around each root and there are 6 sets with each shear, as illustrated in Fig.~\ref{fig:tree}.

\subsection{Image simulations}
\label{appendix:Image_simulations}

The galaxy images are created according to the \textit{forward process} using a Monte Carlo simulation technique. The general idea is that the intensity of a pixel in the image of a galaxy is directly proportional to the number of photons falling into that pixel. The photon count at each point depends on the intensity distribution (the light profile) of the galaxy. Therefore, if we draw random samples (photons) from the theoretical light profile function and then count the number of photons falling in each pixel, we obtain the image of galaxy with the required light profile. The circular light profile thus obtained is then reshaped by applying the necessary transformations to the coordinates of the photons. Since the point-spread function (PSF) can be considered as a probability distribution, a similar method can be used to simulate it. The light profile of the galaxy is convolved with the PSF and finally pixelized into a {\tt FITS} image.

In general, any Monte-Carlo technique can be used for the simulation of the light profile. We use inverse transform sampling for this purpose. It is conceptually simple and generally applicable
for sampling from a one-dimensional probability distribution.
The basic principle is that, given a continuous random variable $U$ distributed uniformly in $[0,1]$ and a random variable $X$ with cumulative distribution $F$, then $X=F^{-1}(U)$ has distribution $F$.
In other words, to sample from $X$, we generate a random sample $U$ and find the value of $X$ at which the cumulative distribution is equal to $U$.

In order to simulate the photons distributed by a Sersic Law, we need to find the cumulative distribution of the density given by Equation
\ref{eq:sersic_law}. Taking $r_e=1$ and substituting $R=kr^{1/n}$, we obtain the cumulative distribution as
\begin{equation}
F(R)=-\frac{\Gamma(2n,R)}{\Gamma(2n)},
\end{equation}
where $n$ is the Sersic index
and $\Gamma(a,x)$ is the incomplete Gamma function.
The inverse of the distribution can be approximately calculated by using linear interpolation, given that we have an ordered set of values of $\{R,F(R)\}$ for the range of $R$ (e.g., from 0 to 20).

The circular light profile of the galaxy obtained by the method above is made larger, elliptical and rotated according to the values of scale radius $r_e$, axis ratio $q$ and angle $\phi$ respectively. These operations can be represented in the form of matrices as
\begin{equation}
\left(
\begin{array}{c}
x_e\\
y_e\\
\end{array}
\right)
=
\left(
\begin{array}{cc}
\frac{1}{\sqrt{q}\,r_e}\, & 0\\
0 & \,\sqrt{q}\,r_e\\
\end{array}
\right)
\left(
\begin{array}{c}
x\\
y\\
\end{array}
\right),
\label{eq:elliptisize}
\end{equation}
\begin{equation}
\left(
\begin{array}{c}
x_r\\
y_r\\
\end{array}
\right)
=
\left(
\begin{array}{cc}
\cos(\phi)\, & -\sin(\phi)\\
\sin(\phi) & \,\cos(\phi)\\
\end{array}
\right)
\left(
\begin{array}{c}
x_e\\
y_e\\
\end{array}
\right).
\label{eq:rotate}
\end{equation}
The shear from the gravitational lensing is applied next. This operation can be written as
\begin{equation}
\left(
\begin{array}{c}
x_s\\
y_s\\
\end{array}
\right)
=
\left(
\begin{array}{cc}
1+g_1\, & g_2\\
g_2 & 1-g_1\\
\end{array}
\right)
\left(
\begin{array}{c}
x_r\\
y_r\\
\end{array}
\right).
\label{eq:shear}
\end{equation}
For computational simplicity, we combine all of the above operations into a single matrix given by
\begin{equation}
\left(
\begin{array}{cc}
r_e\left((1+g_1) c -g_2 s)\right)/\sqrt{q}\,
& r_e\left(g_2 c -(1-g_1) s \right)/\sqrt{q}\\
r_e\sqrt{q}\left((1+g_1) s +g_2 c \right)
&r_e\sqrt{q}\left(g_2 s +(1-g_1) c \right) \\
\end{array}
\right)
\label{eq:forward_process}
\end{equation}
where $c\equiv \cos \phi$ and $s\equiv \sin \phi$.

Having obtained the light profile of the galaxy, we move on to create a Moffat PSF and convolve it with the galaxy. Using a similar procedure
to that described above for the Sersic profile,
we can simulate Moffat PSF given by the Equation~\ref{eq:moffat}. Each sample from the PSF corresponds to the displacement of the photon when convolved with the galaxy.
The circular galaxy can be scaled to the required FWHM and made elliptical by applying the transformation
\begin{equation}
\left(
\begin{array}{c}
x_p\\
y_p\\
\end{array}
\right)
=
\left(
\begin{array}{cc}
1-e_1\, & -e_2\\
-e_2 & 1+e_1\\
\end{array}
\right)
\left(
\begin{array}{c}
x\\
y\\
\end{array}
\right).
\label{eq:shear}
\end{equation}
Assuming that the number of samples in the light profile and the PSF are the same, the convolution of the image is accomplished
by adding the positions of the galaxy and PSF photons.
The image is pixelized by counting the number of photons falling into each pixel of the postage stamp and then it is normalized.

The galaxy images in GREAT08 contain two different light profiles. The final image is created by adding together two images with different light profiles. If $I_1$ and $I_2$ represent two galaxy images with different light profiles, the final image $I_{\rm final}$ is created by the equation
\begin{equation}
I_{\rm final}=mI_1+(1-m)I_2,
\end{equation}
where $0\leq m \leq1$ is a multiplication factor. Poisson noise is then added to each pixel  according to the SNR.

CCD detectors on ground-based telescopes collect a finite number of photons from both astrophysical objects and atmospheric emission.
We therefore mimic this effect by adding the background level $B=1\times 10^6$ to each pixel, and drawing a number from a Poisson distribution with a mean equal to the total number (background plus galaxy) in each pixel.
For numerical convenience we then subtract $B$ from each pixel.
For the RealNoise simulations, this background is much larger than the contribution from the galaxy, so this process is closely approximated by adding a Gaussian random number of standard deviation $\sqrt{B}$ with zero mean.

Before the noise model is applied, the total flux in the galaxy is set using the SNR given in the catalogue, and the background level discussed above. Details are given in the appendix, but in summary we define SNR as the flux divided by the uncertainty in the flux obtained if the true shape (but not normalisation) of the object is known.

For the purpose of the SNR calculations we approximate the Poisson noise as a Gaussian of standard deviation $\sqrt{B}$ for both LowNoise and RealNoise simulations. We follow the definition
\begin{equation}
{\rm SNR} = \frac{F}{\sigma_F}
\end{equation}
where the flux $F$ is the sum of the galaxy counts in each pixel $I_i$
\begin{equation}
F=\sum_i I_i
\end{equation}
and $\sigma_F$ is the uncertainty in the flux.
In general the uncertainty in the flux depends on the assumptions used to measure it. We make the assumption that the true galaxy shape (profile of counts in all the pixels) is known precisely up to an overall unknown scaling which is proportional to the flux. By considering a $\chi^2$ fit it can then be shown that
\begin{equation}
\sigma_F = \sqrt{B} \frac{F}{\left( \sum_i I_i^2 \right)^{0.5}}
\end{equation}
and therefore the flux can be set such that
\begin{equation}
\left( \sum_i I_i^2 \right)^{0.5} = {\rm SNR} \sqrt{B}.
\end{equation}

We note that the images produced using the above ellipticities and $r_e$ values give some very elliptical images that extend beyond the $39\times39$ postage stamp.

\section{Additional information on methods}
\label{appendix:Additional_information_on_methods}

At the launch of the challenge the GREAT08 Team had put six results on
the leaderboard, accompanied by a code wiki \url{http://great08challenge.pbworks.com} summarising the codes used
and linking to downloadable versions of the code that was used on the
GREAT08 simulations. Over the course of the challenge this wiki was
updated by external GREAT08 participants, several of whom also provided
their codes. The key elements of this code wiki are captured in
Table~\ref{tab:participants_more}.

\begin{table*}
\begin{center}
\begin{tabular}{l@{}ccll}
\hline\hline
\bf{Key} &
\bf{Method name} &
{\bf Language} & {\bf Runtime}  & {\bf URLs}\\
&&
& (s/galaxy)
&\\
\hline\hline
HB &
CVN Fourier  &
Matlab & 1 &
\url{http://great08challenge.pbworks.com/f/Great-Challenge.zip}
\\ \hline
AL &
CLT KK99 &
f90 & 0.4 &
\url{http://cosmologist.info/utils/StackedShear.zip}
\\ \hline
TK &
Lensfit &
C & 0.08 & \url{http://www.physics.ox.ac.uk/lensfit/}
\\
&&&&\url{http://www-astro.physics.ox.ac.uk/~tdk/files/lensfit.tar.gz}\\
\hline
CH &
KSBf90 &
f90 & 0.005 & \url{http://www.roe.ac.uk/~heymans/KSBf90}
\\ \hline
PG &
gfit  &
Python & 0.2 &
\\ \hline
MV &
KKshapelets with flexion &
f77, f95 & 0.03 &
\\ \hline
KK       &
KKshapelets &
f77 & 0.03 &
\url{http://www.strw.leidenuniv.nl/~kuijken/shear-shapelets.html}
\\\hline
HHS &
Gauss and variants &
Python & 0.05 &
 \\ \hline
SB &
im2shape &
C & 0.02 &
\url{http://www.sarahbridle.net/im2shape/great08_im2shape_v1.0.tar.gz}
\\ \hline
MJ    &
BJ02 deconvolved shapelets  &
C++ & 0.08 &
\url{http://www.hep.upenn.edu/~mjarvis/great08/v1.tar.gz}
\\ \hline
USQM &
Unweighted stacked quadrupole moments &
Matlab & 0.001 &
\url{www.sarahbridle.net/usqm_v1.0.tar.gz}
\\ \hline
\hline
\end{tabular}
\end{center}
\caption{
Table providing more details about the methods. This table collates
information from the GREAT08 code wiki \url{http://great08challenge.pbworks.com} on the programming language
used, an indicative time taken per galaxy, and associated URLs.
These runtimes are only illustrative since they are reproduced as provided by the code authors and no attempt has been made to benchmark or compare the machines used. The TK method takes 0.01 seconds using 8 threads, and these numbers are multiplied to give the number in the table, for ease of comparison with other methods.
\label{tab:participants_more}}
\end{table*}

\section{Detailed analysis of results}
\label{appendix:Detailed_analysis_of_results}

\subsection{LowNoise$\_$Blind}
\label{appendix:Detailed_analysis_of_results:LNB}

\begin{figure*}
\center
\epsfig{file=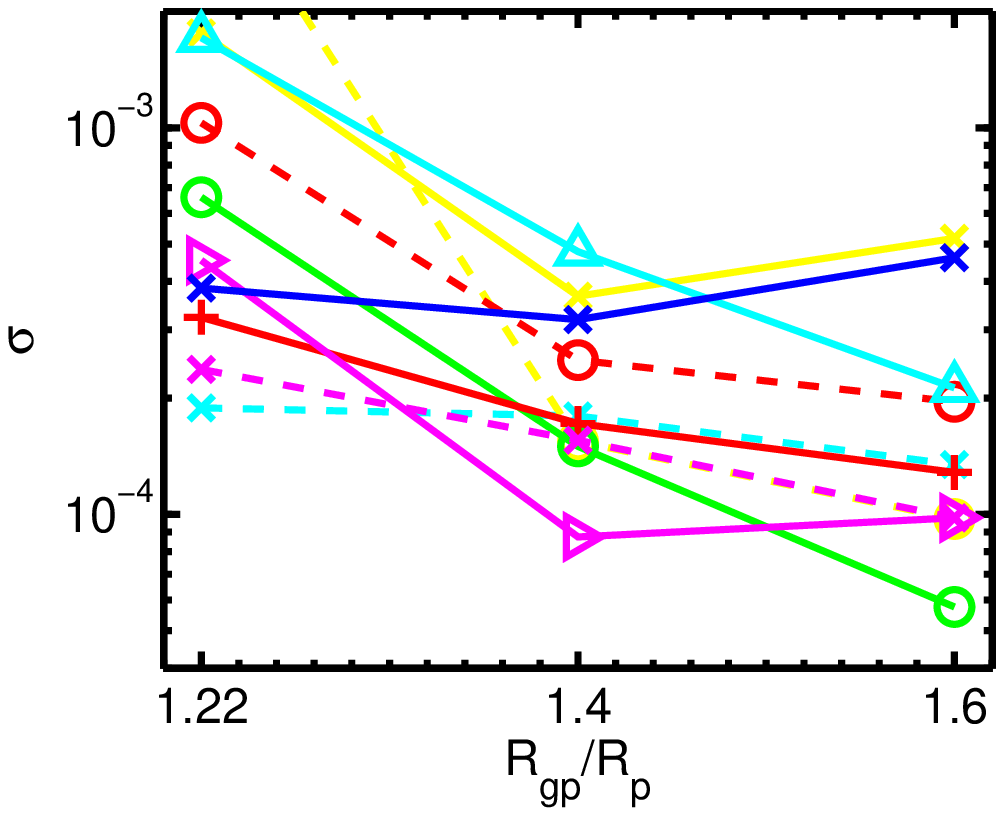,height=5cm,angle=0}
\epsfig{file=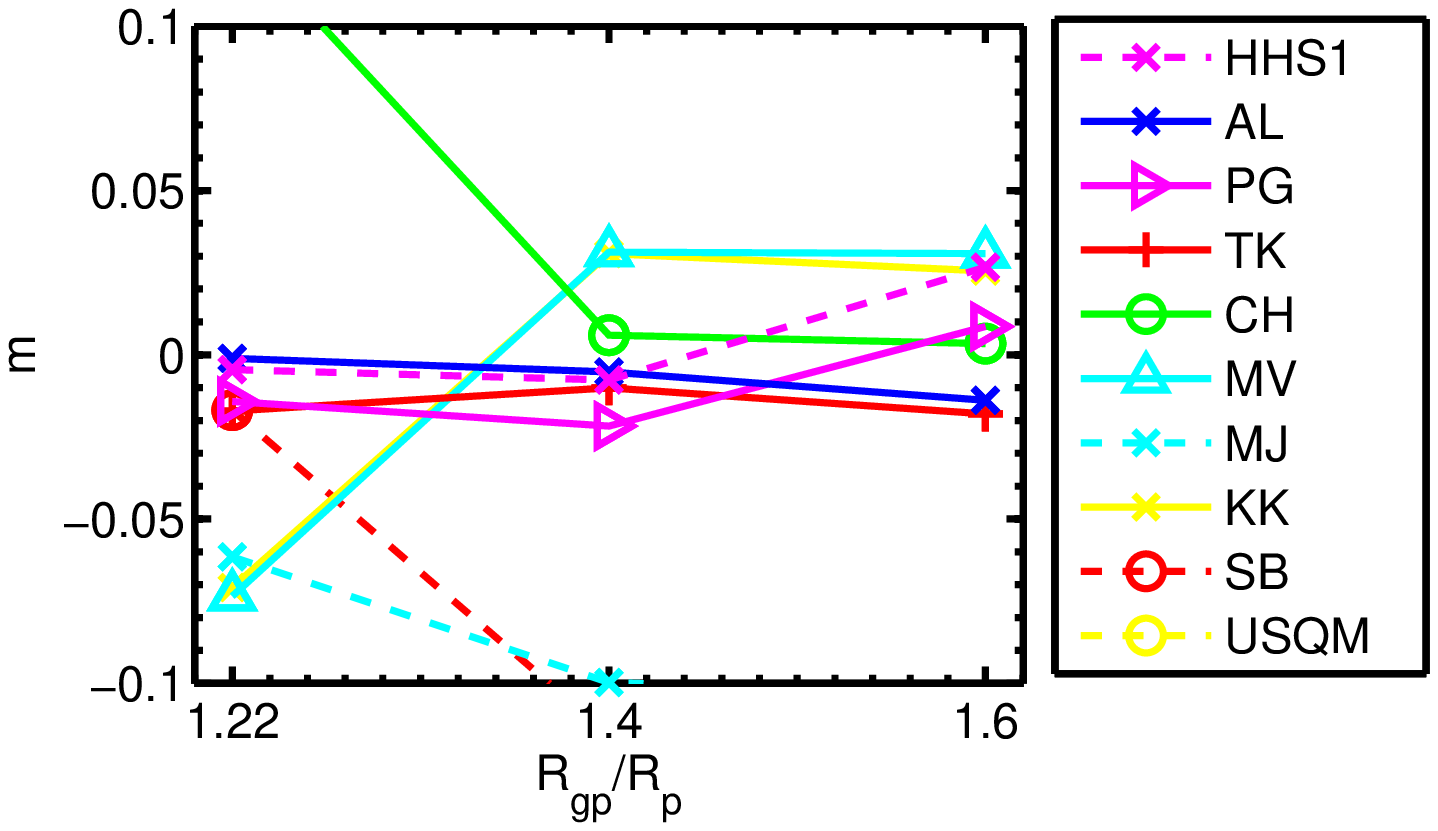,height=5cm,angle=0}\\
\epsfig{file=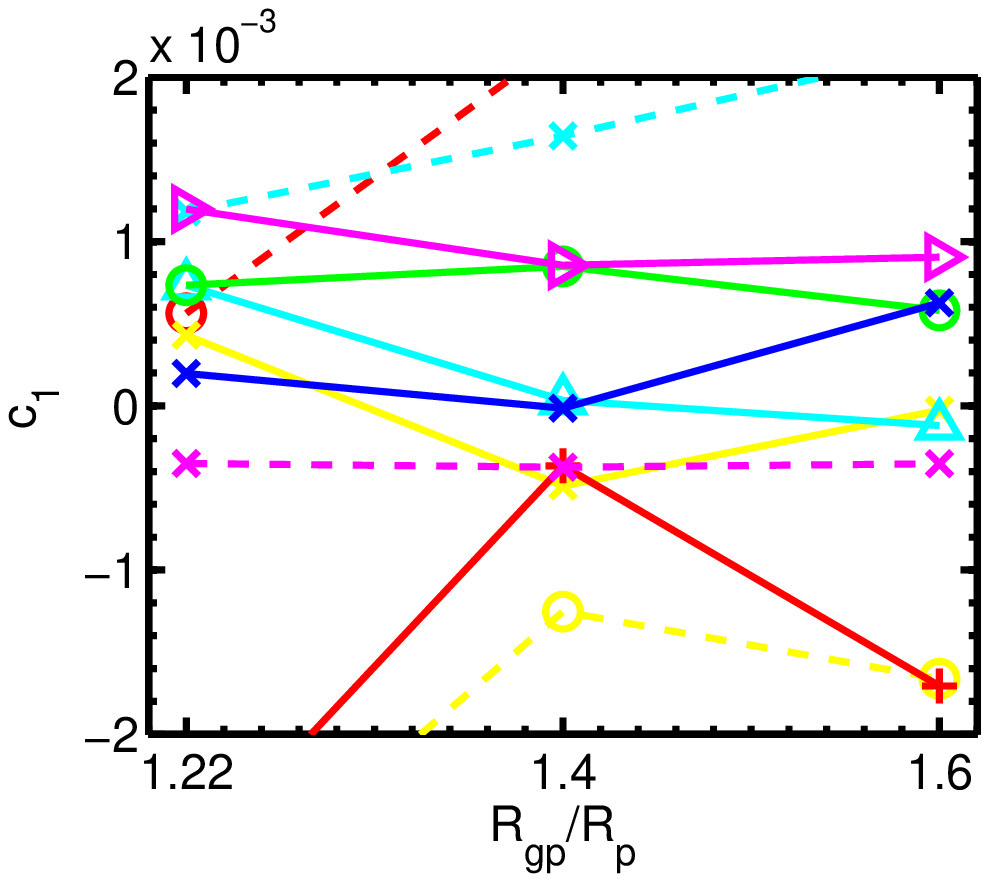,height=5.3cm,angle=0}
\epsfig{file=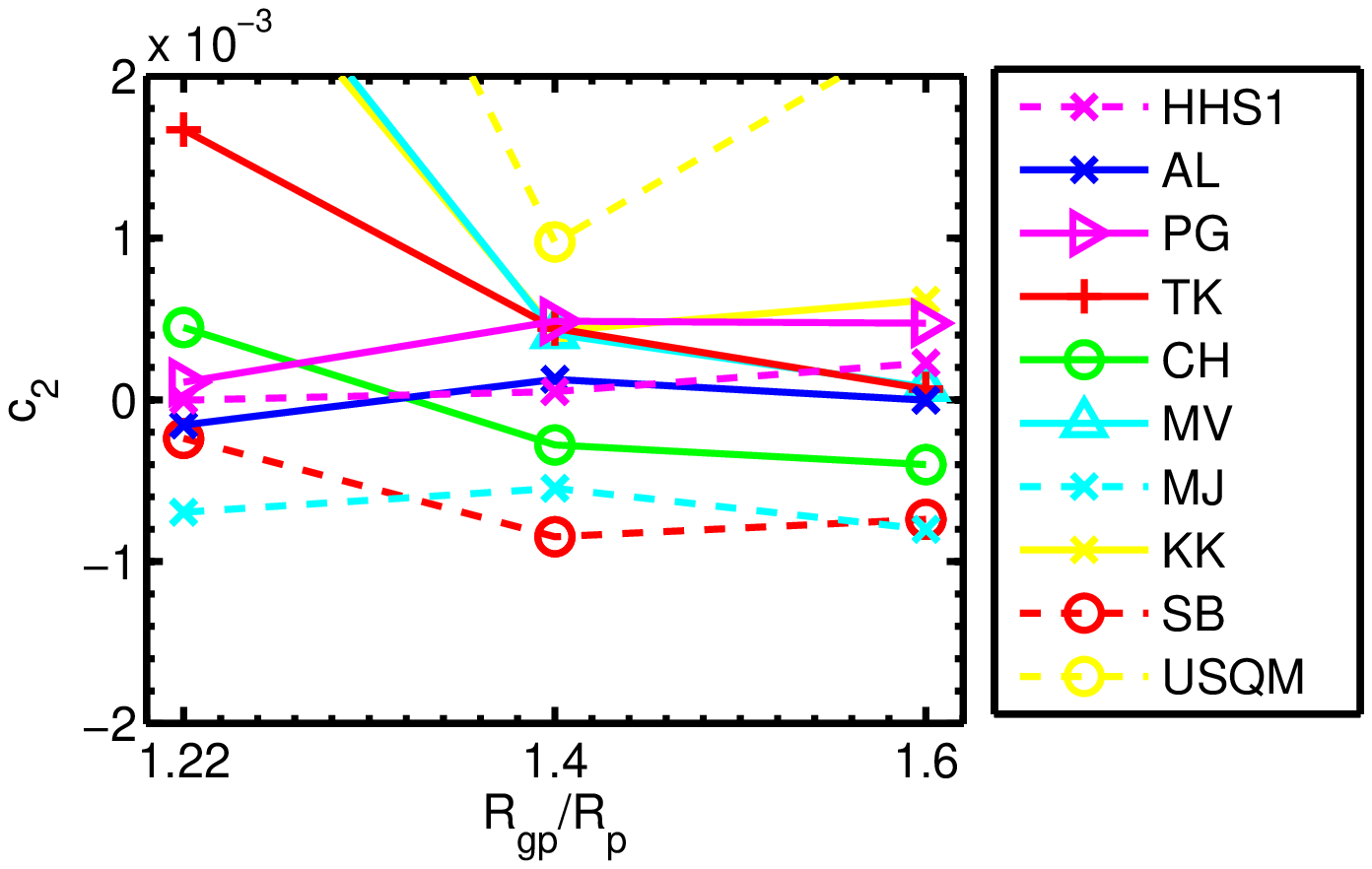,height=5.3cm,angle=0}
\caption{
Scatter, multiplicative and additive shear measurement bias  as a function of galaxy size for LowNoise\_Blind.
}
\label{fig:mc_breakdowns_LNB}
\end{figure*}

The overall performance, as measured by $Q$, has contributions from various competing effects. We break these up into a multiplicative bias $m$, an additive bias $c$ and an rms dispersion $\sigma$, as defined in Section~\ref{sect:Requirements}.
For each of the three simulation branches in LowNoise$\_$Blind we fit a straight line to a plot of submitted $g_1$ versus true $g_1$ values and identify the slope as $(m_1+1)$ and $c_1$ as the offset. We repeat for $g_2$
and average the multiplicative biases together to obtain an overall value $m$, similarly for the additive bias $c$.
The scatter $\sigma$ is given by the standard deviation of the residuals. Note that although the 90 degree rotations in GREAT08 substantially reduce the effect of shape noise, this would be a large additional contribution to the statistical uncertainty from realistic data, as it roughly adds in quadrature with the statistical scatter (at the level of about 0.2 per galaxy).

The finite number of simulations means that these values cannot be determined exactly. Therefore we also estimate uncertainties on the fitted multiplicative and additive biases from the submitted shear values.
The uncertainty on $m$ depends on the shear measurement method used and on the simulation properties. We calculate the uncertainty on the estimated $m_i$ by calculating the likelihood as a function of $m_i$ and $c_i$ and marginalising over $c_i$. We then calculate an average uncertainty on $m$ over shear components $i$. Uncertainty decreased with increasing galaxy size for most methods, and the winning method HHS1 had one of the smaller uncertainties on $m$, decreasing from $5\times10^{-3}$ at $R_{gp}/R_p=1.22$ to $2.3\times10^{-3}$ at $R_{gp}/R_p=1.6$.
This may be compared to the multiplicative bias values $m$ obtained by different groups, and we see that the uncertainty is small compared to at least one of the values obtained by each group and therefore is not the limiting factor in interpreting these results.

The uncertainties on the additive biases $c_1$ and $c_2$ also decrease with increasing galaxy size, as expected. At a given galaxy size they range over almost an order of magnitude for the different methods. A typically low uncertainty was obtained by HHS1 across the range of galaxy sizes, and it varies from $10^{-4}$ at  $R_{gp}/R_p=1.22$ to $3\times10^{-5}$ at $R_{gp}/R_p=1.6$. Again, this is much smaller than the additive shear biases seen by all groups for at least one galaxy size and is therefore not the limiting factor in obtaining small biases.

Fig.~\ref{fig:mc_breakdowns_LNB} shows the multiplicative bias $m$ and additive bias $c$ as a function of
${R_{gp}/R_p}$ for LowNoise\_Blind.
We now see that
HHS1, performs less well at large galaxy sizes due to an increased multiplicative bias, indicating that the shears are overestimated for these galaxy sizes.
In the $Q$ plot (Fig.~\ref{fig:Q_breakdowns_LNB}) the second highest method, AL (blue solid line), does best at the fiducial size and worse at larger and smaller sizes. On the more detailed figures of multiplicative and additive biases we see that the picture seems yet more curious, with good $m$ and $c$ values (close to zero) at small galaxy sizes, and becoming worse at large sizes. A more detailed analysis
shows that the slight improvement at the fiducial galaxy size can be attributed to a partial cancelation between the effects of a negative $m$ and a positive $c$.
The third best result, PG, is relatively insensitive to the galaxy size; this effect is mirrored in the additive bias, which dominates the overall $Q$ result
since the multiplicative bias is relatively small.

We see that the CH method acquired a very large
positive $m$ at small ${R_{gp}/ R_p}$, indicating a consistent $\sim
12\%$ overestimation of the true shear when the galaxies are poorly
resolved.
TK has best performance all round on the fiducial model
and this may be expected because LensFit was optimised to work well on typical galaxies used for cosmic shear, which therefore tends to coincide with the fiducial model used for GREAT08.
The fact that MJ, SB and USQM consistently underestimate the shear is the dominant contribution to their poor performance.
MV and KK both underestimate the shear at small
${R_{gp}/ R_p}$, but overestimate the shear at moderate and large
${R_{gp}/ R_p}$.  Note that the MV method is an extension of the KK
method, and the two performed very similarly in all the
LowNoise\_Blind plots.
Several methods (KK, MV, USQM, CH) had the largest additive biases $c$ for poorly resolved galaxies,
which may suggest that the information about the true PSF model was not fully incorporated into their analyses.

As discussed in Section~\ref{sect:Requirements}, a successful method needs to produce reasonably low noise shear measurements, which we quantify by the scatter $\sigma$,
shown in Fig.~\ref{fig:mc_breakdowns_LNB}.
The scatter decreases as the galaxy size is increased, which is expected as information on the galaxy can be obtained from more image pixels. There is about an order of magnitude difference between the methods, with HHS1 having a consistently low scatter around $10^{-4}$.
Since there are 10,000 galaxies in each {\tt FITS} file this corresponds to an uncertainty on the shear of each individual galaxy of 0.01, which is typical for a SNR of 200.
For LowNoise\_Known there is only a single {\tt FITS} file for each simulation branch, which means that there is no sum over files $j$ in Eq.~\ref{eq:Q} (i.e. $j=1$).  So, in the absence of other biases ($m=c=0$) we would have $Q_l \sim 10^{-4} / \sigma_k^2$, where $\sigma_k$ is the scatter for a single simulation branch.
Therefore $\sigma_k<3\times10^{-4}$ is required to reach the target of $Q_l\sim1000$ for a given simulation branch.
Some methods have $\sigma_k\sim10^{-3}$ at the smallest galaxy sizes, which will limit their overall $Q$ to around 100.

\subsection{RealNoise$\_$Blind}
\label{appendix:Detailed_analysis_of_results:RNB}

\begin{figure*}
\center
\epsfig{file=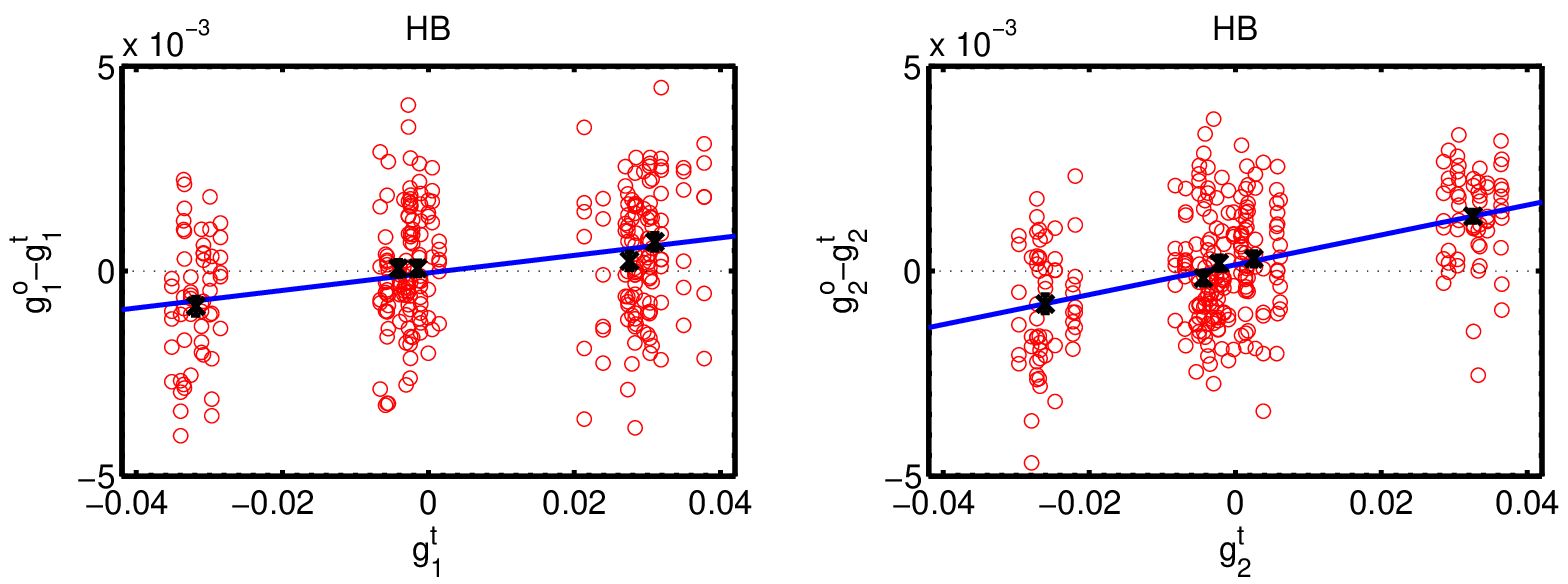,width=17cm,angle=0}
\epsfig{file=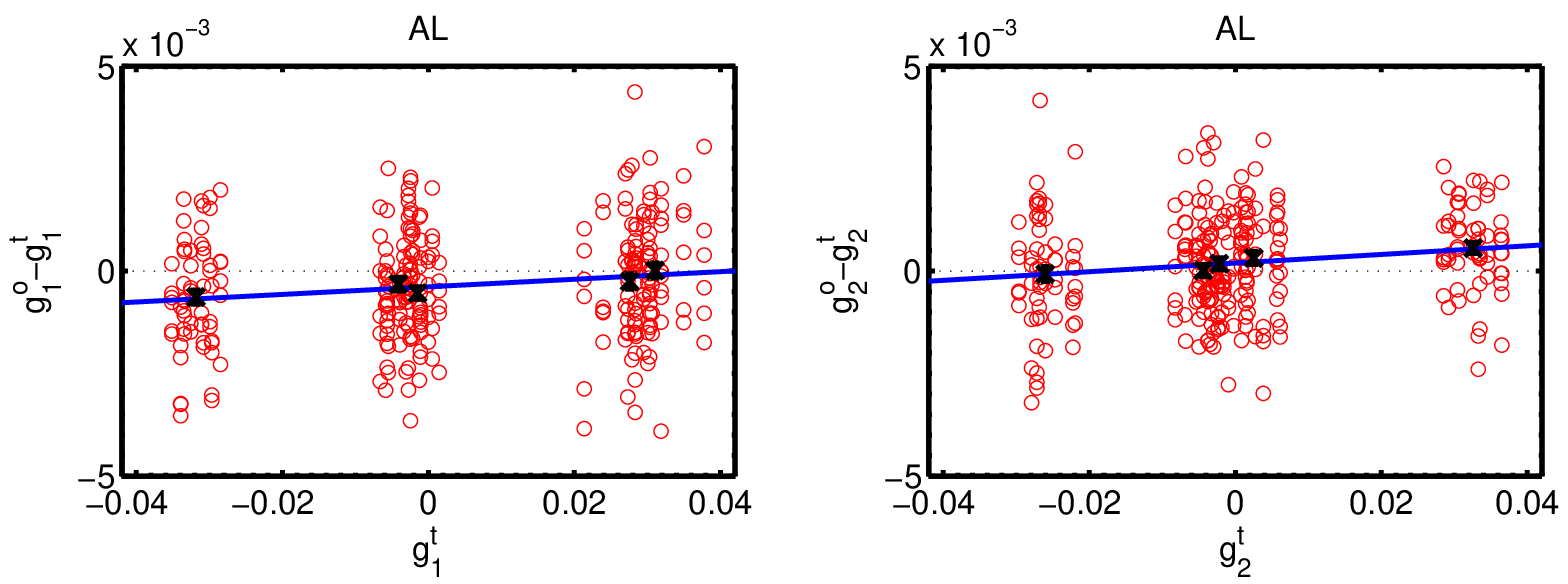,width=17cm,angle=0}
\caption{
Residual shear versus true shear for shear component 1 (left panels) and shear component 2 (right panels) for the top two methods HB (upper panels) and AL (lower panels).
The circles show results for each set in the fiducial simulation branch and the crosses and error bars show the average residual for each root shear value for the fiducial branch.
}
\label{fig:go_gt_RNB}
\end{figure*}

Fig.~\ref{fig:go_gt_RNB} shows the output shear residuals versus the input
shear for the top two methods, for the fiducial simulation branch. This figure illustrates how the multiplicative and additive errors are calculated.  The total $Q$ for a given simulation branch is roughly a combination of the slopes and offsets of each best-fit line, and the scatter about the lines.
The equivalent point for LowNoise$\_$Blind has only five points on it, and the circles are identical to the crosses.

\begin{figure*}
\center
\epsfig{file=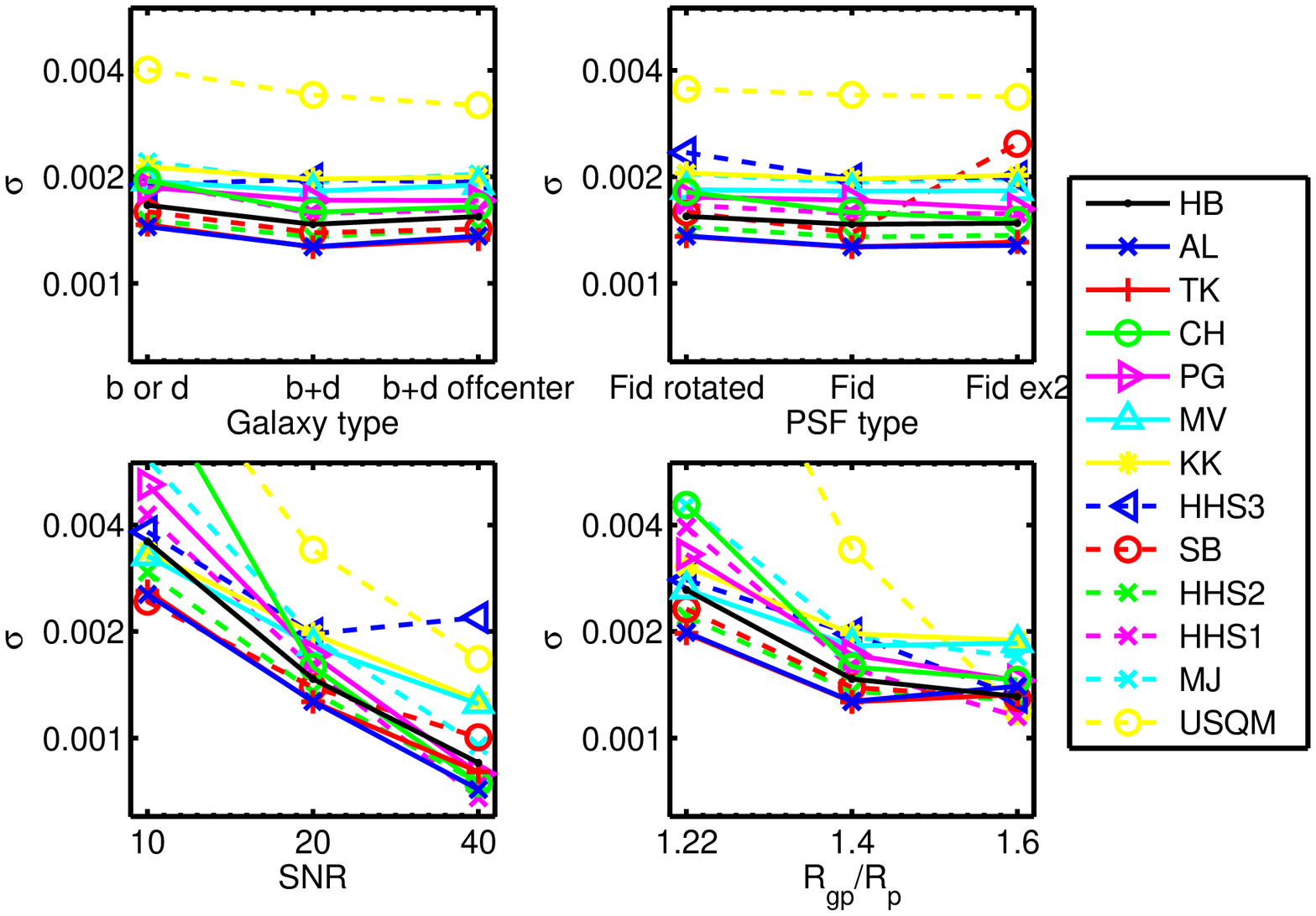,height=11.5cm,angle=0}
\epsfig{file=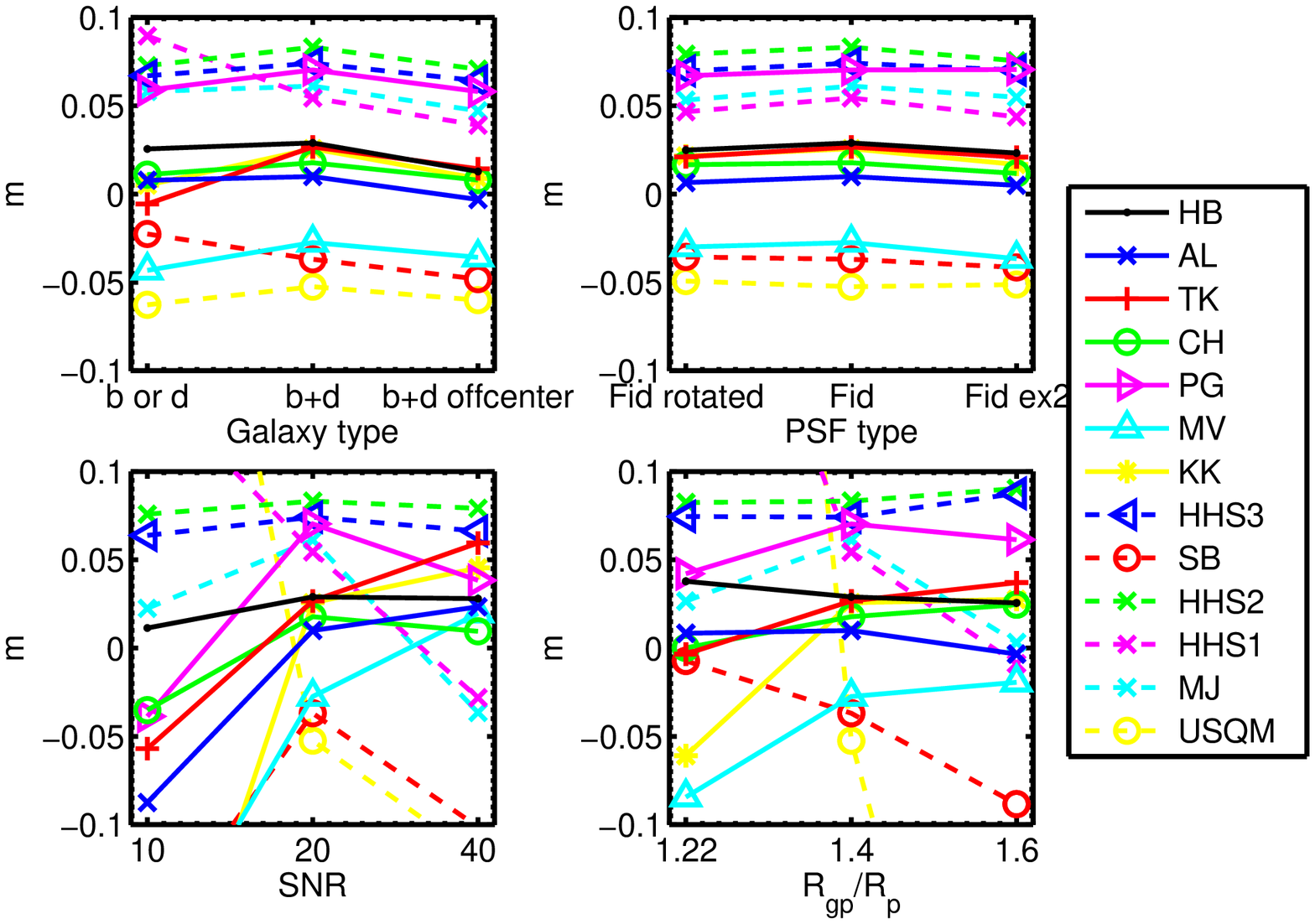,height=11.5cm,angle=0}
\caption{
Upper panel: Scatter about a linear fit to output versus input shear.
Lower panel: multiplicative shear measurement bias  as a function of galaxy size for RealNoise\_Blind.
}
\label{fig:sigmam_breakdowns_RNB}
\end{figure*}

\begin{figure*}
\center
\epsfig{file=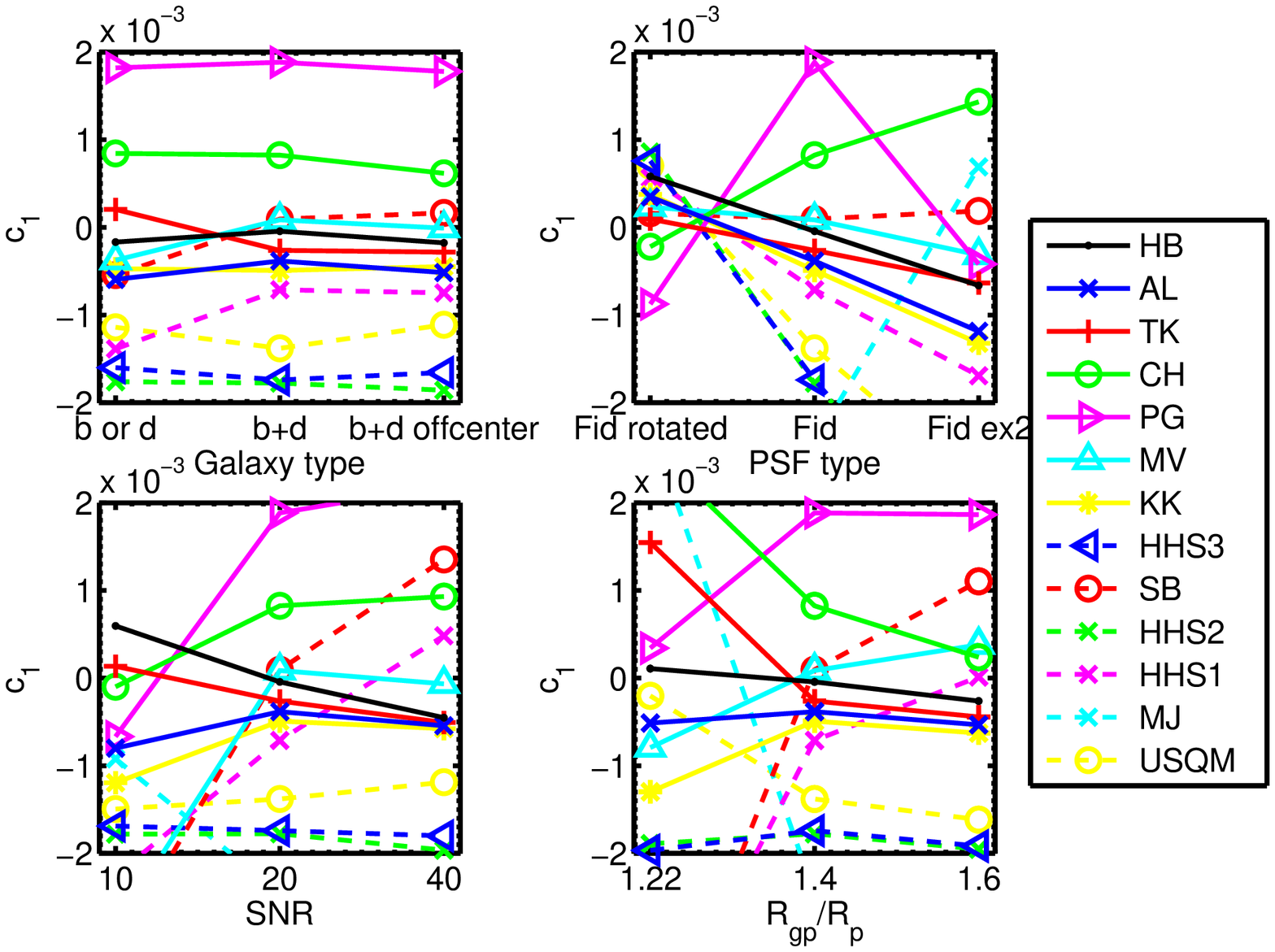,height=11.5cm,angle=0}
\epsfig{file=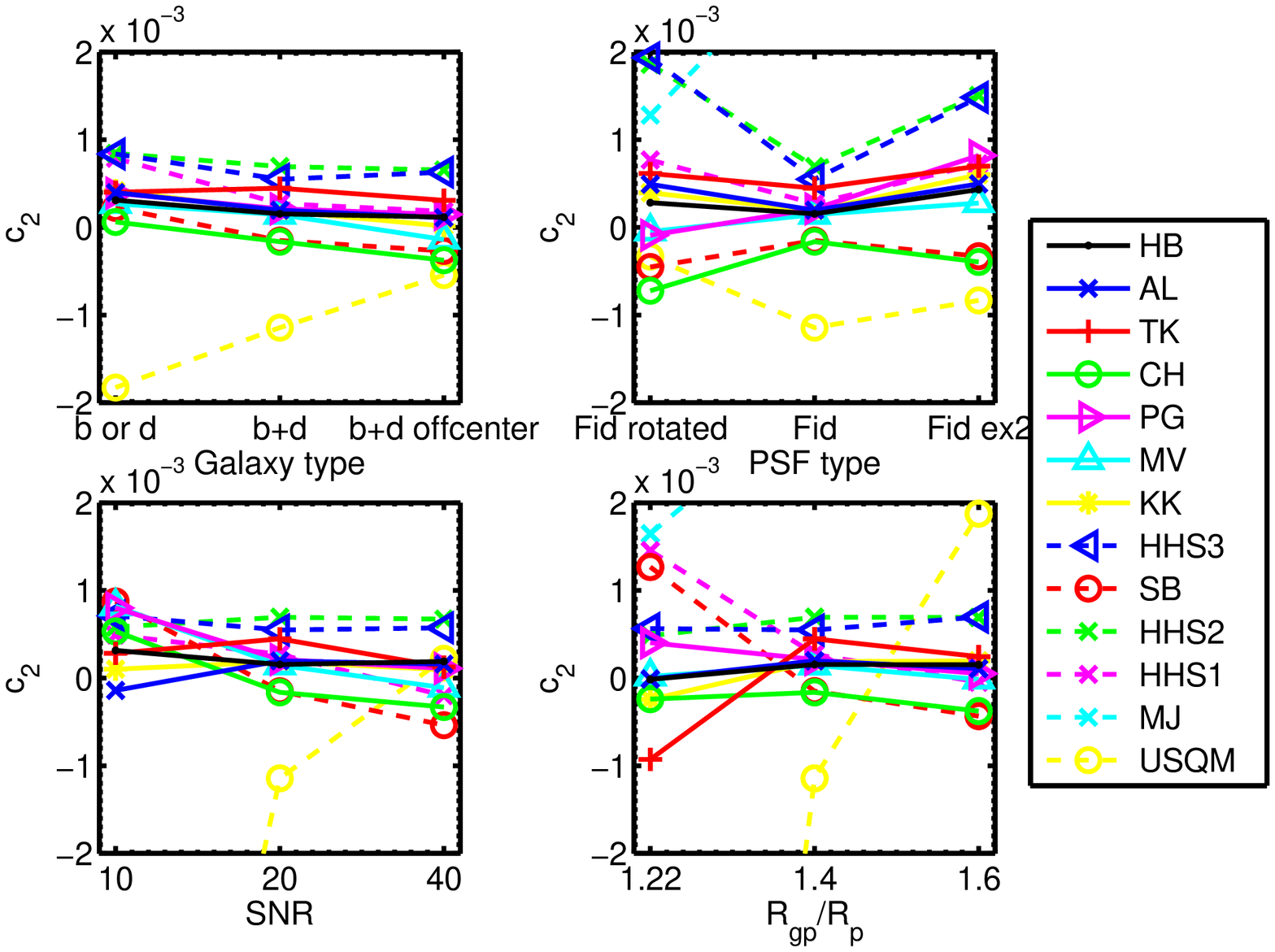,height=11.5cm,angle=0}
\caption{
Additive shear measurement bias  as a function of galaxy size for RealNoise\_Blind.
}
\label{fig:c_breakdowns_RNB}
\end{figure*}

We show the multiplicative and additive biases in Figs.~\ref{fig:sigmam_breakdowns_RNB} and~\ref{fig:c_breakdowns_RNB}.
The decreased SNR in RealNoise\_Blind is compensated for by averaging over many shear values to reduce the noise and ensure that the quality measures $Q$, $m$ and $c$ can be dominated by systematic biases.

We first consider overall trends in multiplicative and additive biases
(Figs.~\ref{fig:sigmam_breakdowns_RNB} and \ref{fig:c_breakdowns_RNB}).
The
``psftype'' panels indicate that changes in the PSF had virtually no
effect on $m$ but quite a large effect on $c$.
Incorrect estimation of the PSF size tends to cause a multiplicative bias, so given that the PSFs all had roughly the same size, and varied only in ellipticity,
this result is not surprising.
There is a general tendency for $c$ to be best for the fiducial PSF, positive for the ``PSF rot'' and negative for ``PSF $e \times$ 2''.
This tendency indicates that the participants made the most efforts to model the fiducial PSF, which is used for almost all of the simulations.
It would be interesting to compare the observed trend with the result of wrongly assuming the fiducial PSF for the two other PSF branches in case this explains the result.

The scatter of the submitted shears about the best fit line can be seen qualitatively by the range of the circles in Fig.~\ref{fig:go_gt_RNB}, and quantitatively for each simulation branch in Fig.~\ref{fig:sigmam_breakdowns_RNB}.
Typical values around $10^{-3}$ are averaged down in the $Q$ calculation in the average over $j=1,...,300$ simulations in a given simulation branch which have similar shear values. Therefore $Q_l \sim 300 \times 10^{-4} /\sigma_k^2$, and $\sigma_k$ should be less than about $5\times10^{-3}$ for all simulation branches to prevent a method with $m=c=0$ from reaching $Q\sim1000$.
This condition is met by most methods even at the lowest SNR value.

The uncertainties on the multiplicative bias are close to constant with respect to galaxy and PSF type and decrease with increasing SNR and galaxy size.
With the exception of USQM, there is little scatter between the groups for a given simulation branch (tens of percent difference), and the smallest uncertainties are obtained by AL and TK.
For these methods, since the uncertainty on $m$ is always less than $10^{-2}$, 
we infer that the finite number of simulations is not the dominant reason that every submission departs from zero multiplicative bias for at least one simulation branch.
The uncertainties on the additive bias are always less than $2\times10^{-4}$ for the best methods and therefore also do not dominate the biggest departures from perfection.

For the method HB,
the multiplicative calibration bias (upper panels, Fig.~\ref{fig:sigmam_breakdowns_RNB}) is very close to constant with simulation branch.
The shears are consistently overestimated by about 2 per cent.
This bias is above our detailed simplistic requirements for far future experiments, but note that if a method really did have a multiplicative bias that was completely constant with the properties of the simulation or universe, then it would be trivially removed by dividing all shears by the relevant number.
The additive calibration bias for this method is always below our detailed requirement of $0.0003$ for far future experiments, except for the ``Fid rotated'' PSF branch and the low SNR branch. It would be intriguing to know if this could be fixed further by more detailed modeling of the PSF.

The poor performance of AL on the low SNR branch
appears to come mostly from a multiplicative bias of nearly 10\% (Fig.~\ref{fig:Q_breakdowns_RNB}).
The results on the most elliptical PSF (``Fid $e \times$ 2'') are also relatively disappointing, and come from the large additive calibration bias
(Fig.~\ref{fig:Q_breakdowns_RNB}). This result is consistent with a problem with modeling this particular PSF, in which residual PSF ellipticity remains to add to the true shear.

The good results of MV at high SNR and large galaxy size
is largely due to the reduction in multiplicative bias in these regimes. This result could possibly hint at inaccurate modeling of the PSF size.

The poorer performance for smaller galaxy sizes for HHS now seems to come from both an increased multiplicative and additive error.
The multiplicative bias increases slightly as a function of galaxy size in LowNoise\_Blind but decreases as a function of galaxy size in RealNoise\_Blind.
Perhaps there is some kind of cancelation between the increasingly negative multiplicative bias as a function of SNR and the large positive multiplicative bias seen in RealNoise\_Blind at smaller galaxy sizes.
HHS2 and HHS3 used stacking to decrease dependence on the assumed galaxy
model.
The additive calibration bias is still significant and reduces the overall $Q$ value. The sharp changes in additive calibration bias with PSF type suggest that
the PSF is not being sufficiently well modelled.

In STEP2 there was found to be a systematic difference between $m_1$ and $m_2$ that was attributed to the different effective pixel scales in the two directions. We have made separate figures for $m_1$ and $m_2$ but find them to be visually similar for most methods except CH.
Galaxy type variations in general had little effect on $m$ and $c$ overall, a surprising result also found in STEP2.

\end{document}

%% file: great08figure.tex



\graphicspath{{.}}  




\begin{center}
\begin{tikzpicture}[node distance=3cm]
	\tikzstyle{every entity}=[draw=black!100,thick]

  \node[entity] (d) at (5,0) {{Shear (${g}$)}} ;

  \node[entity] (c) at (0,0) {Ellipticities}
  edge [-stealth, very thick]  node[auto,swap] {Averaging} (d);

  \node (b) at (5,5) {
  \includegraphics[width=0.07\textwidth]{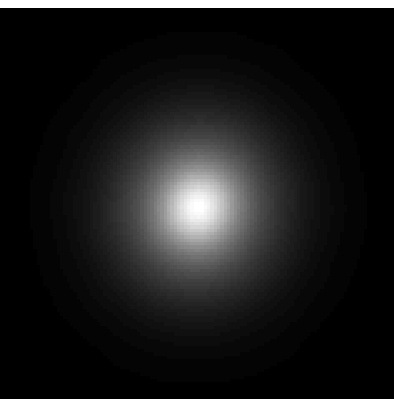}}          			 edge [-stealth, very thick]  node[right,swap] 
  {\begin{minipage}[t]{3cm} 
  Model fitting\\{\small (e.g. spline)} 
  \end{minipage}} 
  (d);

  \node (a) at (0,5)  {\includegraphics[width=0.1\textwidth]{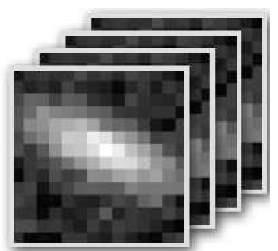}}
  edge [-stealth, very thick] node[above,swap] 
  {\begin{minipage}[t]{2.4cm} 
  Stacking {\small (e.g. in Fourier domain)} 
  \end{minipage}}  
  (b)
  edge [-stealth, very thick] node[left,swap] 
  {\begin{minipage}[t]{2cm} 
  Model fitting\\ {\small (e.g. shapelets)} 
  \end{minipage}} 
  (c);

\end{tikzpicture}
\end{center}
